\documentclass[11pt]{article}
\usepackage[utf8]{inputenc}
\usepackage{authblk}
\usepackage{dirtytalk, float}
\usepackage[top=.8in, bottom=0.75in]{geometry}
\usepackage{epsfig, color}
\usepackage{amsfonts, amsmath, setspace}
\usepackage{mathtools, array, booktabs}
\usepackage{amssymb, cite, soul, url}
\usepackage[normalem]{ulem}
\usepackage{scalerel}
\usepackage{graphicx, float, caption, subcaption}
\usepackage{lineno}

\linenumbers

\usepackage{verbatim}
\usepackage{appendix}
\usepackage{amsmath}
\usepackage{graphicx}
\usepackage{lineno}
\usepackage{array}
\usepackage{longtable}
\usepackage{natbib}
\usepackage{svg}
\setcitestyle{aysep={}}
\linenumbers
\usepackage{subcaption}
\usepackage{tabularx}
\usepackage[colorlinks,allcolors=black,pdftex]{hyperref}

\title{Impact of Forest Canopy Structure on Buoyant Plume Dynamics During Wildland Fires
}

\author[1]{Antonio Quim Cervantes}
\author[1,3]{Ajinkya Desai}
\author[2,3]{Tirtha Banerjee}

\affil[1]{Department of Civil and Environmental Engineering, University of California, Irvine, CA 92697}
\affil[2]{Department of Earth System Science, University of California, Irvine, CA 92697}
\affil[3]{Lawrence Livermore National Laboratory, Livermore, CA 94550}

\date{October 2025}

\begin{document}
\nolinenumbers 
\maketitle

\begin{abstract}
This study investigates the influence of forest canopy heterogeneity on buoyant plume dynamics resulting from surface thermal anomalies representing wildland fires, utilizing Large Eddy Simulation (LES). The Parallelized Large-Eddy Simulation Model (PALM) was employed to simulate six canopy configurations: no canopy, homogeneous canopy, external plume-edge canopy, internal plume-edge canopy, 100 m gap canopy, and 200 m gap canopy. Each configuration was analyzed with and without a static surface heat flux patch of 5000 $\mathrm{W \cdot m^{-2}}$, resulting in a resting buoyant plume. Simulations were conducted under three crosswind speeds: 0, 5, and 10 $\mathrm{m \cdot s^{-1}}$. Results show that canopy structure significantly modifies plume behavior, mean flow, and turbulent kinetic energy (TKE) budgets. Plume updraft speed and tilt varied with canopy configuration and crosswind speed. Pressure gradients associated with plume updrafts were modified based on the canopy configuration, resulting in varying crosswind speed reductions at the plume region. Strong momentum absorption was observed above the canopy for the crosswind cases, with the greatest enhancement in the gap canopies. Momentum injection from below the canopy due to the heat source was also observed, resulting in plume structure modulation based on canopy configuration. TKE was found to be the largest in the gap canopy configurations. TKE budget analysis revealed that buoyant production dominated over shear production. At the center of the heat patch, the gap canopy configurations showed enhanced buoyancy within the gap. Spatial distributions demonstrated increased shear production at the interface between the plume and crosswind.


\end{abstract}

\section{Introduction}
\label{intro}

Wildfire spread rate generally depends on three factors: fuel (makeup, size, configuration, moisture), weather, and terrain \citep{curry_forest-fire_1939}. Mechanisms for fire spread include convective heat transfer, radiative heat transfer, and fire spotting \citep{tohidi_experimental_2016}, which is intricately linked with the turbulence associated with fire-vegetation-atmosphere interaction \citep{heilman2023atmospheric}. The interaction between fire and the atmosphere significantly impacts fire spread mechanisms \citep{desai_features_2023}. For example, \cite{cervantes_numerical_2023} demonstrated that the background turbulence intensity influenced simulated ember travel distances \citep{chung_interaction_2023, petersen2024characterizing}. \citet{banerjee2020effects} and \citet{banerjee_impacts_2020} demonstrated that alterations of the nature of fire-vegetation-atmosphere interaction can modify the efficacy of fuel management practices such as forest thinning. With growing concern about more destructive wildfires and a proliferation of prescribed fire implementation, how fires interact with heterogeneities of vegetation fuel requires further investigation. Abrupt changes in vegetation canopy structure, such as horizontal transitions from a grassland to a forest (edges) or a clearing within the canopy (gaps), have been shown to impact ambient atmospheric turbulence \citep{banerjee_mean_2013, dupont_coherent_2009}. The presence of a fire within this configuration further complicates this phenomenon \citep{kiefer_study_2016}. 

Before addressing fire-atmosphere interactions in heterogeneous canopies, it is helpful to first discuss canopy flows in the absence of fire plumes. The forest canopy acts as a distributed momentum sink, where wind velocity is reduced in the canopy layer due to drag forces imposed by the plant elements (leaves, branches, etc.). This creates a roughness layer at the canopy top where shear production is high, and turbulent kinetic energy (TKE) is maximum \citep{poggi2004effect,dias-junior_large_2015}. Above the canopy, an inflection point for the mean streamwise velocity is associated with the formation of shear-induced Kelvin-Helmholtz instabilities, and the canopy sub-layer flow is characterized as a mixing layer as opposed to a canonical boundary layer \citep{brunet_turbulent_2020, finnigan_turbulence_2000}.

When the canopy features heterogeneities in the form of gaps or edges, pressure gradients are created from the abrupt change in spatial distribution of vegetation along the mean wind direction \citep{banerjee_mean_2013, dupont_coherent_2009}. The upstream flow from a clearing to the windward side of a canopy edge experiences an increase in the dynamic pressure perturbations, and behind the leading edge, the pressure decreases until it settles to a steady state further downstream \citep{banerjee_mean_2013}. As a response, the streamwise velocity is reduced, and the vertical velocity increases for a short distance downstream due to deflection by the canopy, and flow recirculations are formed \citep{kroniger_flow_2018}. \cite{dupont_coherent_2009} also describes the turbulence behavior of edge flows as a multi-stage process. At the edge interface, there are ejection-dominant fluctuations that form Kelvin-Helmholtz instabilities, and an internal boundary layer starts to develop. Downstream, this ejection dominance starts to transition into sweep dominance, which triggers vorticity at the canopy top. Further along, these structures break down and create more complex shapes, such as hairpin vortices. Additionally, a distinct shear layer develops, which contributes significantly to the generation of TKE \citep{dupont_coherent_2009}. At the leeward edge of a plant canopy, flow separation occurs due to a low-pressure region, which results in a recirculation zone \citep{banerjee_mean_2013}. This is followed by a re-attachment zone further downstream, resembling similar characteristics of a back-facing step flow \citep{fontan_flume_2013, detto_structure_2008}. 

The combination of these effects at the windward and leeward sides of an edge canopy is seen in gaps. Gaps perturb the mean flow and create pressure gradients, with low pressure at the upstream edge and high pressure at the downstream edge \citep{banerjee_mean_2013}. This results in an acceleration of wind speed within the gap, and a deceleration as wind flows over the downwind edge of the gap and into the canopy \citep{banerjee_mean_2013, pimont_validation_2009, schlegel_large-eddy_2015, fontan_flume_2013}. Higher wind speeds from aloft enter the gap and form recirculations \citep{banerjee_mean_2013, kiefer_study_2016, schlegel_large-eddy_2015}. Lab-scale experiments \citep{fontan_flume_2013} and field-scale measurements \citep{schlegel_large-eddy_2015} show that TKE is greater within canopy gaps than within the canopy itself. These experimental studies also show that momentum reduction is largest in the gap. \cite{fontan_flume_2013} further reports that for a gap size ratio of $L/H = 1$, where the gap length ($L$) equals the canopy height ($H$), the downwind edge constrains the upstream shear layer, therefore limiting eddy size growth. For a gap ratio of 3, the shear layer fills the gap but does not fully reattach to the floor, whereas for a gap ratio of 5, reattachment occurs before the downwind edge \citep{fontan_flume_2013}.

Research on fire-atmosphere interactions in heterogeneous forest canopies remains limited; however, there are more studies on fire plumes in no-canopy situations as well as homogeneous canopies. In quiescent conditions with no canopy present, plumes from buoyancy point sources rise vertically with axial symmetry \citep{raupach_similarity_1990}. In crossflow, plumes may become tilted in the direction of flow depending on the buoyancy strength and wind speed \citep{devenish_entrainment_2010}. With the canopy present, plume behavior is shaped by canopy-induced turbulence, though buoyancy remains the dominant influence \citep{raupach_similarity_1990}. \cite{desai_features_2023} analyzed field-scale measurements of sub-canopy and grassland fires, finding that the near-surface buoyant production of a grassland fire was dominated by shear production; however, buoyancy increased with height. For sub-canopy fires \citep{desai_features_2023}, shear and buoyancy production were more pronounced at the canopy top, but at mid-canopy height, buoyancy was more dominant. \citet{desai2024investigating} investigated the organization of momentum and sensible heat flux within the canopy volume using wavelet analysis of tower based experimental data on sub-canopy heading fires. They identified a strong degree of organization of the turbulence field in terms of warm updrafts and cold downdrafts. Depending on the shape and strength of the buoyant plume structure, the degree of organization was found to vary, and fire induced coherent structures were found to remain coherent over shorter periods of time compared to no fire conditions, especially for vertical momentum flux events, compared to sensible heat flux. In addition, the tilting of the fire plume by the sweeping eddies causes fire-induced perturbations in the temperature and velocity fields to occur earlier at the top of the canopy, before the fire front reaches the canopy base. Moreover, oscillatory motions of fire plumes have been observed to be impacted by the level of canopy turbulence \citep{chung_interaction_2023}. However, the fundamental impact of canopy structural heterogeneity on plume structure and energetics remains poorly understood.

In heterogeneous canopies, fire plumes near edges encounter stronger horizontal winds that increase plume tilt. In contrast, plumes within canopies rise more vertically until reaching the canopy top, where they are deflected by crosswinds \citep{meroney_fires_2007}. \cite{kiefer_study_2016} investigated buoyant plumes in a no-canopy, gap canopy, and homogeneous canopy domain using numerical modeling. It was found that the presence of the gap and the plume located in the center of the gap had the largest impact on mean flow and TKE compared to other configurations.  

In this paper, the effect of canopy heterogeneity and wind forcing on buoyant plume dynamics are explored. In particular, edge and gap flow dynamics in the presence of a buoyant plume are compared against homogeneous and no-canopy plume scenarios. Through the use of a large eddy simulation model, this study addresses how canopy edges and gaps modify plume structure under varying wind conditions, and the role of canopy heterogeneity on the mean flow and turbulence that govern plume rise and spread. The structure of the paper details the methodology and experimental campaign, results, and discussion.

\section{Methodology}
\label{sec:1}
\subsection{Large Eddy Simulation Model}
To investigate complex canopy-plume interactions, the Parallelized Large Eddy Model (PALM) is employed. This computational fluid dynamics solver is optimized for high-performance computing using 2-D domain decomposition \citep{maronga_parallelized_2015}. PALM’s Large Eddy Simulation (LES) model solves the non-hydrostatic incompressible Navier-Stokes equations in Boussinesq approximated form. A sink term is applied to the momentum equation $(- c_DLAD\sqrt{u^2_i}{u_i})$ when using PALM's built-in plant canopy module, which describes the drag forces created by the plant canopy. The LES model resolves the discretized filtered equations of turbulence to the grid scale. Sugrid-scales are modeled by the 1.5-order Deardorff closure scheme \citep{maronga_overview_2020}.

The forest canopy is represented as a volume-averaged porous medium. The leaf area per unit volume of the tree foliage is parameterized by the leaf area density (LAD) profile (seen in Fig. \ref{fig:general_setup}). The LAD profile is based on \cite{dias-junior_large_2015}, and is incorporated as a sink term in the momentum equation. This LAD profile is chosen since the simulations are well validated by \citet{dias-junior_large_2015}; however, substantial differences in the outcome are not expected, and different LAD profiles could easily be used instead.

\subsection{Experimental Campaign \& Model Setup}
\begin{figure}[ht!]
    \centering
    \begin{subfigure}{\textwidth}
            \caption{General Setup}
            \includegraphics[width=\linewidth]{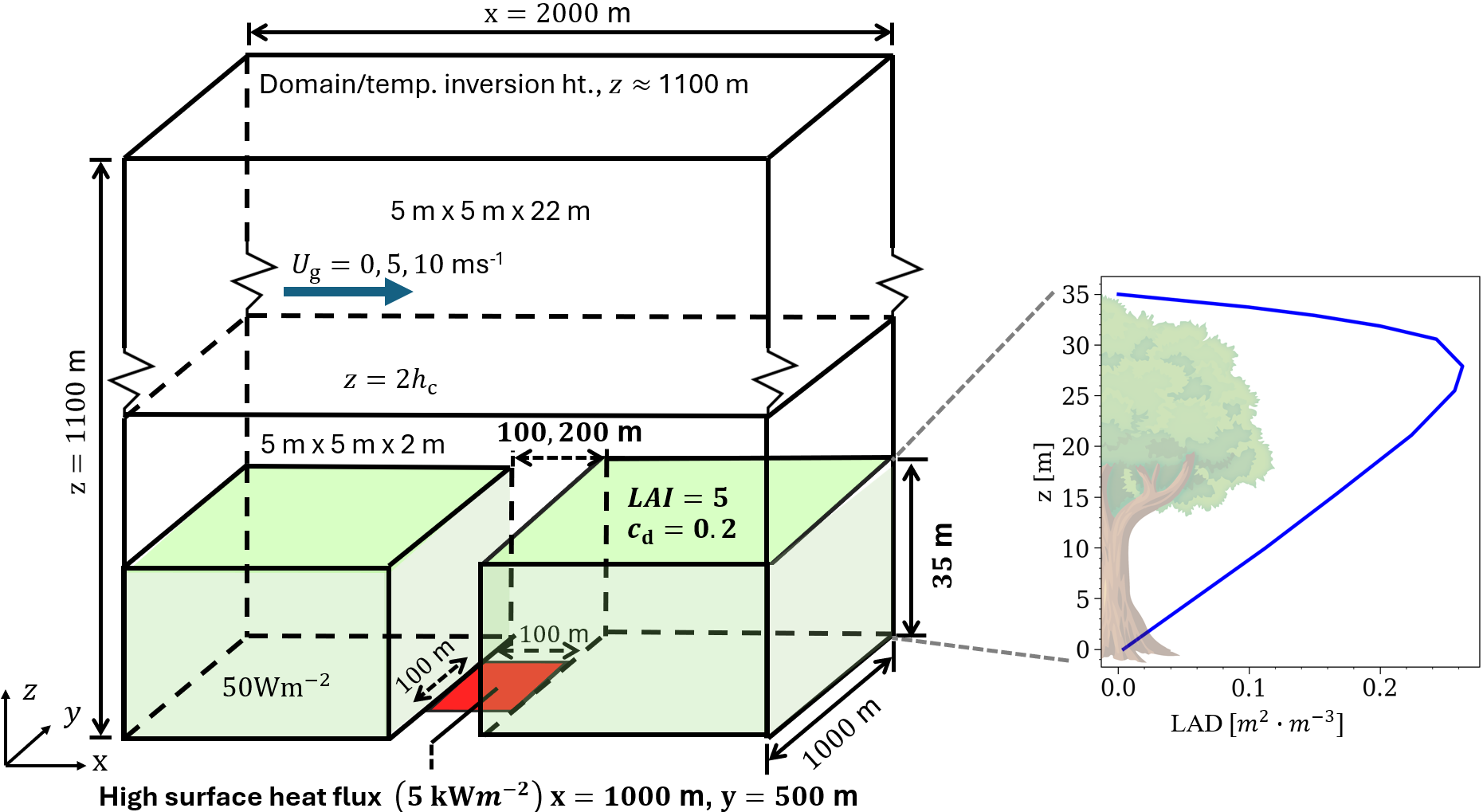}
            \label{fig:general_setup}
    \end{subfigure}

    \begin{subfigure}{\textwidth}
        \caption{Canopy Configurations}
        \includegraphics[width=\linewidth]{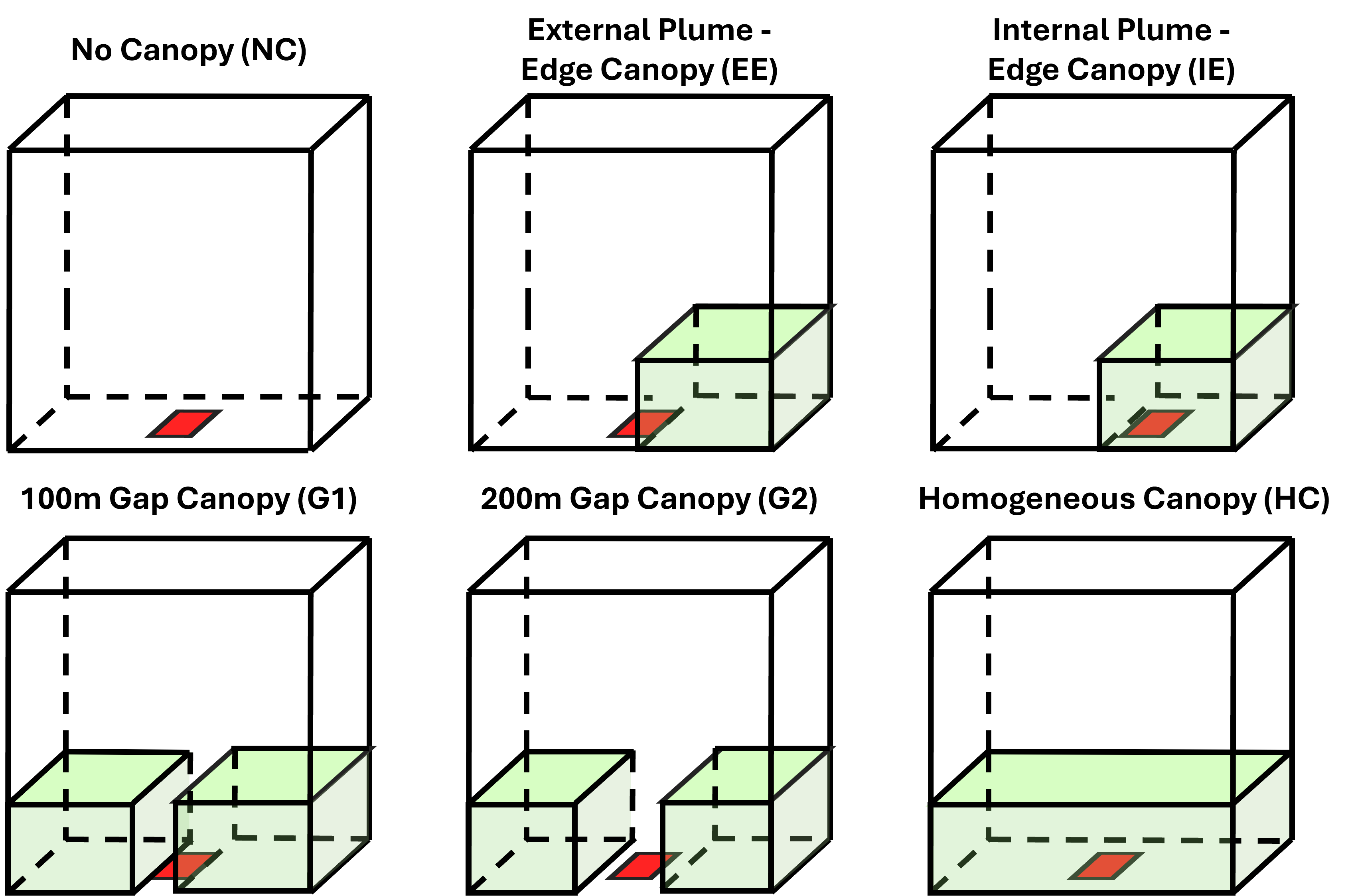}
        \label{fig:canopy_configuration}
    \end{subfigure}
     \caption{(a) Schematic of LES general setup (left) and leaf area density profile used in the simulations (right). (b) Canopy configurations from the experimental campaign.}
     \label{fig:setup}
\end{figure}

Six forest canopy configurations were simulated to capture the effects of canopy heterogeneity, plume location, and plume-atmosphere interactions (see Fig. \ref{fig:canopy_configuration}). As outlined in Table \ref{tab:simulation_campaign}, each configuration was tested with three prescribed geostrophic wind speeds of 0, 5, and 10 $\mathrm{m\cdot s^{-1}}$. These speeds were chosen to assess the plume's behavior under varying atmospheric conditions. All cases were performed with and without a plume generated by a 5000 $\mathrm{W\cdot m^{-2}}$ surface heat flux over a 100 × 100 m square patch. When the plume was not present, the heat patch was set to the surrounding surface heat flux of 50 $\mathrm{W\cdot m^{-2}}$, which was applied to all the runs, to represent typical daytime conditions. The heat patch location was fixed at the center of the domain, and did not move or expand over time. For the EE and IE cases, the patch was positioned externally or internally to the edge of the canopy, respectively. To maintain a temperature inversion at the top of the boundary layer, a negative surface heat flux of –10 $\mathrm{W\cdot m^{-2}}$ was applied at the domain's upper boundary. 

All simulations shared domain dimensions of 2000 m (length), 1100 m (height), and 1000 m (width), except for edge canopy cases (length 1500 m). Grid spacing throughout the entire domain was 5 m in the $x$ and $y$ direction, and below 70 m height, grid spacing in the $z$ direction was 2 m. Above 70 m, grid stretching was applied in the z direction with a stretch factor of 1.08, up to a maximum grid height of 22 m. The canopy height ($h$) was set at 35 m, with a drag coefficient of 0.2 and a leaf area index (LAI) of 5. The LAD profile was uniformly applied under the green-shaded areas in Fig. \ref{fig:canopy_configuration}. The gap size of G1 and G2 were equal to and twice the heat patch size, respectively.

Table \ref{tab:simulation_campaign} summarizes boundary conditions: non-cyclic inflow/outflow to avoid upstream contamination by the plume, except for the no-wind cases which used cyclic inflow/outflow, cyclic lateral boundaries, free-slip top, and no-slip bottom (with 0.02 m roughness). Atmospheric conditions were neutral, and each simulation ran for 10 hours, both with and without the plume.

% Simulation Campaign Table
\begin{table}[ht!]
    \centering
\caption{Simulation Campaign, Boundary and Initial Conditions}
\label{tab:simulation_campaign}
    \begin{tabular}{|l|l|}\hline
    \multicolumn{2}{|c|}{Canopy Configurations}\\\hline %\hline
         Name & Acronym \\ \hline
         no canopy & NC \\ \hline 
         external plume-edge canopy & EE\\ \hline
         internal plume-edge canopy & IE\\ \hline
         100 m gap canopy & G1\\ \hline
         200 m gap canopy & G2\\ \hline
         homogeneous canopy & HC\\ \hline \hline
         
    \multicolumn{2}{|c|}{Domain Settings}\\\hline %\hline 
         Grid Size (dx, dy) & 5 m, 5 m \\ \hline 
         Grid Size (dz, below $z=70$ m ) & 2 m \\ \hline 
         Grid stretch factor (above $z=70$ m ) & 1.08 \\ \hline 
         Max dz (above $z=70$ m ) & 22 m\\ \hline
         Domain length, width, height & 2000 m, 1000 m, 1100 m\\ \hline 
         Grid points & 399, 199, 199 \\ \hline \hline
    
    \multicolumn{2}{|c|}{Boundary Conditions}\\\hline %\hline 
         Inlet/Outlet & Dirchelet/Radiation\\ \hline 
         North/South Walls & Cyclic\\ \hline 
         Top/Bottom Walls & No-Slip/Free-Slip\\ \hline
         Surface Roughness Length & 0.02 m\\ \hline \hline

    \multicolumn{2}{|c|}{Initial Conditions}\\\hline %\hline
        Geostrophic Wind Speeds & 0, 5, 10 $\mathrm{m \cdot s^{-1}}$ \\\hline
        Plume surface heat flux & 50, 5000 $\mathrm{W \cdot m^{-2}}$ \\\hline
        Atmospheric Stability & Neutral \\ \hline
        Run time & 10 hrs\\ \hline 
        Canopy Height & 35 m\\ \hline
        Canopy Drag Coefficient & 0.2 \\ \hline
    \end{tabular} 
\end{table}

\subsection{Model Validation}
\begin{figure}[ht!]
    \centering
    \includegraphics[width=0.5\textwidth]{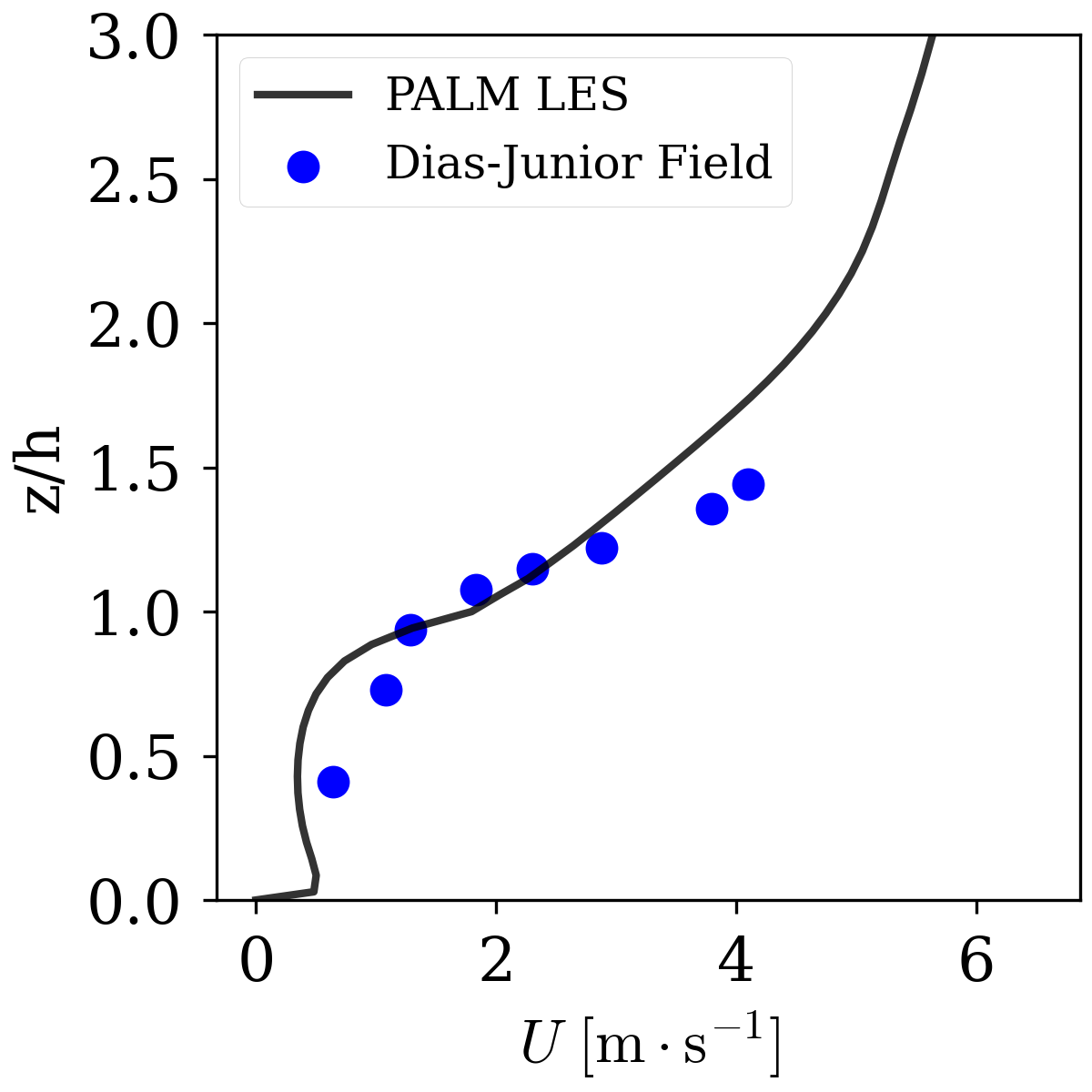}
    \caption{Validation of LES model with field results from \cite{dias-junior_large_2015}. Velocity profile obtained from the center of the domain averaged over one hour of simulation time.}
    \label{fig:validation}
\end{figure}

The PALM model was validated against field data collected in the Amazon during the  Large-Scale Biosphere-Atmosphere Experiment in Amazonia (LBA) campaign, digitzed from \cite{dias-junior_large_2015}. The homogeneous canopy configuration at 10 $\mathrm{m \cdot s^{-1}}$ wind speed without the buoyant plume was used for validation. The model's horizontal velocity data from the last hour of simulation time was averaged and plotted against the field data, as shown in Fig. \ref{fig:validation}. The model's velocity profile closely matched the field measurements, capturing both the velocity reduction below the canopy the logarithmic increase of the velocity above it.

\section{Results and Discussion}

\subsection{Overview of Simulation Cases}
The influence of canopy structure on buoyant plume dynamics was evaluated using streamwise cross-sections taken at the domain center from the simulated flow fields. Mean and turbulence statistics were computed from masked outputs, which output velocity data at every timestep of the simulation. Because PALM uses a variable timestep, the average velocity output was recorded at 0.3-second intervals. This data was taken from the final hour of each 10-hour simulation. In the following sections, the effects of plume presence, wind speed, and canopy configuration are examined.

\subsection{Mean Flow Statistics}

\subsubsection{Baseline Canopy Flow (No Plume)}

\begin{figure}[htbp]
    \centering
    \renewcommand{\thesubfigure}{\roman{subfigure}} % makes the subfigure captions as lowercase roman numerals
    \begin{subfigure}{\textwidth}
            % \caption{$U = 5 \ \mathrm{m \cdot s^{-1}}$}
            \includegraphics[width=\linewidth]{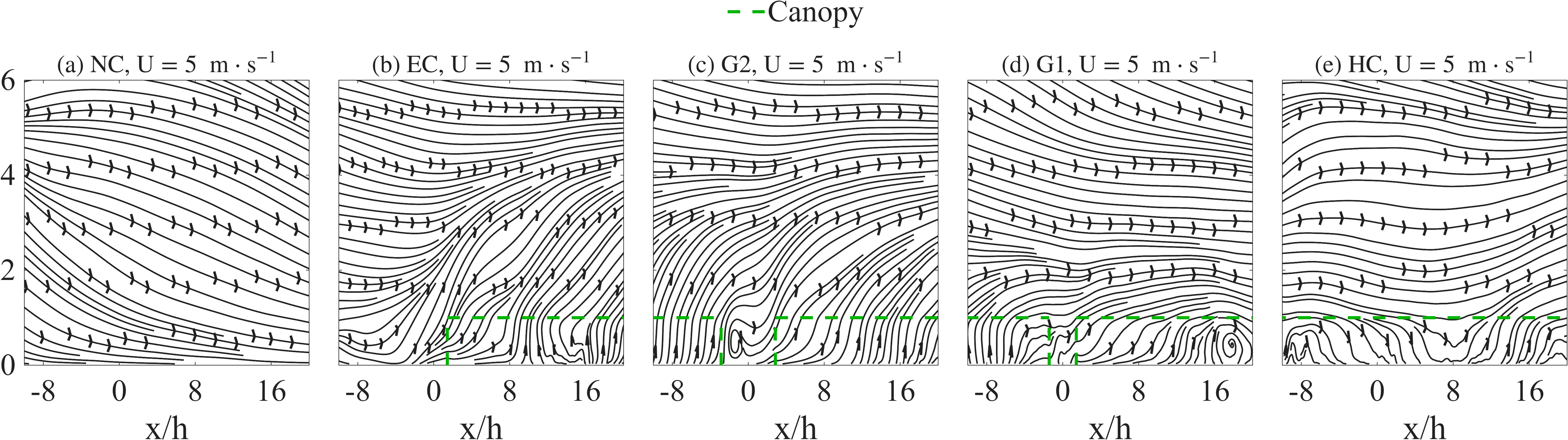}
            \label{fig:U5 no plume streamlines}
    \end{subfigure}

    \begin{subfigure}{\textwidth}
        % \caption{$U = 10 \ \mathrm{m \cdot s^{-1}}$}
        \includegraphics[width=\linewidth]{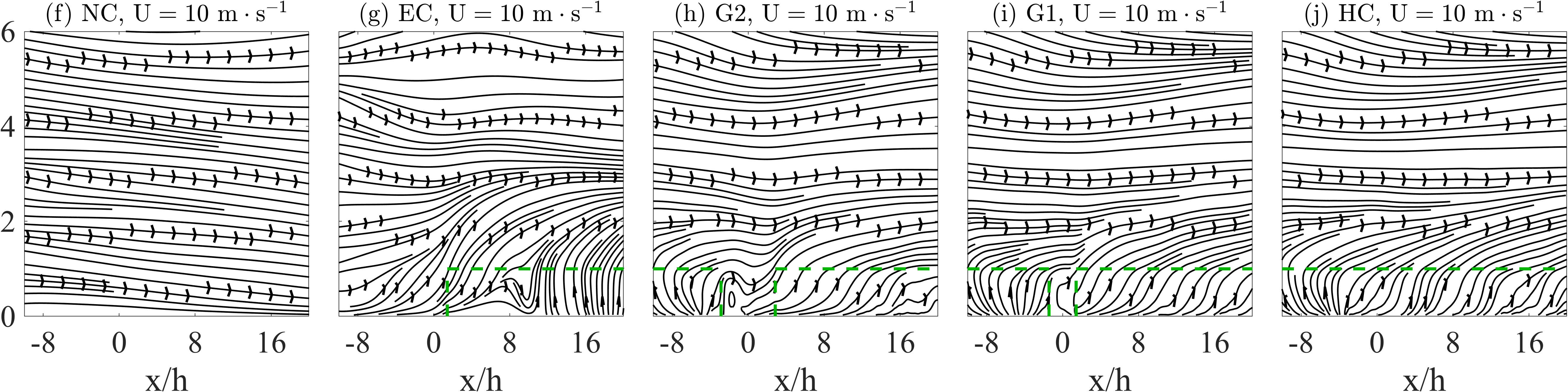}
        \label{fig:U10 no plume streamlines}
    \end{subfigure}
     \caption{Velocity streamlines in the XZ plane at $y=500 \ \mathrm{m}$ for the no plume crosswind cases $\mathrm{U=5 \ m \cdot s^{-1}}$ (top row) and $\mathrm{U=10 \ m \cdot s^{-1}}$ (bottom row).}
    \label{fig:No plume streamlines}
\end{figure}

% intro
Before examining the plume in crosswind cases, it is important to assess the qualitative features of the canopy flows with no plume (no heat patch). Figure \ref{fig:No plume streamlines} shows the streamlines of the no plume cases at 5 and 10 $\mathrm{m \cdot s^{-1}}$.

% Edge canopy: IBL development, recirculations zones
The edge canopy, EC (same canopy configuration as EE and IE), shows the development of an internal boundary layer (IBL) above the canopy (Fig. \ref{fig:No plume streamlines}), similar to the LES findings of \cite{dupont_coherent_2009} and \cite{banerjee_mean_2013}. The 5 $\mathrm{m \cdot s^{-1}}$ wind speed displays a taller IBL height and steeper updrafts generated by the edge than the 10 $\mathrm{m \cdot s^{-1}}$ case. Recirculation zones are found in the EC cases at $x/h = 14$ for 5 $\mathrm{m \cdot s^{-1}}$, and $x/h = 6$ for 10 $\mathrm{m \cdot s^{-1}}$. It is interesting to note that in other studies such as \cite{dupont_coherent_2009} and \cite{banerjee_mean_2013}, recirculation zones were observed at about x/h = 5 downstream from the edge. However in our study, the impact of wind speed in modifying the location and extent of the recirculation zone is evident. 

% Gap case windspeed difference
For the gap cases, a recirculation zone is observed in the 100 m and 200 m gap cases just after the upstream edge of the gap. In the G1 case, the streamlines in Fig. \ref{fig:No plume streamlines} do not clearly show this feature; however, the flow reversal is evident in Fig. \ref{fig:U5 u no plume} in the appendix. A secondary recirculation zone also develops in the 100 m gap case at 5 $\mathrm{m \cdot s^{-1}}$ near $x/h = 18$. At 10 $\mathrm{m \cdot s^{-1}}$ wind speed, the G2 case shows a more compact recirculation zone constrained to the high-speed flow above the gap, while the far-downstream recirculation zone seen in the G1 case at 5 $\mathrm{m \cdot s^{-1}}$ is no longer present.
 
% HC windspeed differences
The streamlines for the NC cases at both wind speeds indicate relatively uniform mean flow. The HC case at 5 $\mathrm{m \cdot s^{-1}}$ displays some downdrafts followed by updrafts within the canopy. At 10 $\mathrm{m \cdot s^{-1}}$, there are persistent updrafts.

\subsubsection{Plume in No Crosswind $(0 \ \mathrm{m \cdot s^{-1}})$}

\begin{figure}[htbp]
    \centering
    \begin{subfigure}{0.25\textwidth}
        % \caption{}
        \includegraphics[width=\linewidth]{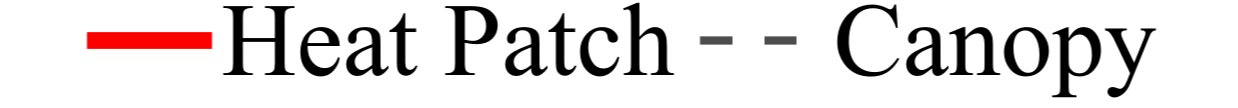}
    \end{subfigure}
    \hfill
    \begin{subfigure}{\textwidth}
        % \caption{}
         \includegraphics[width=\textwidth]{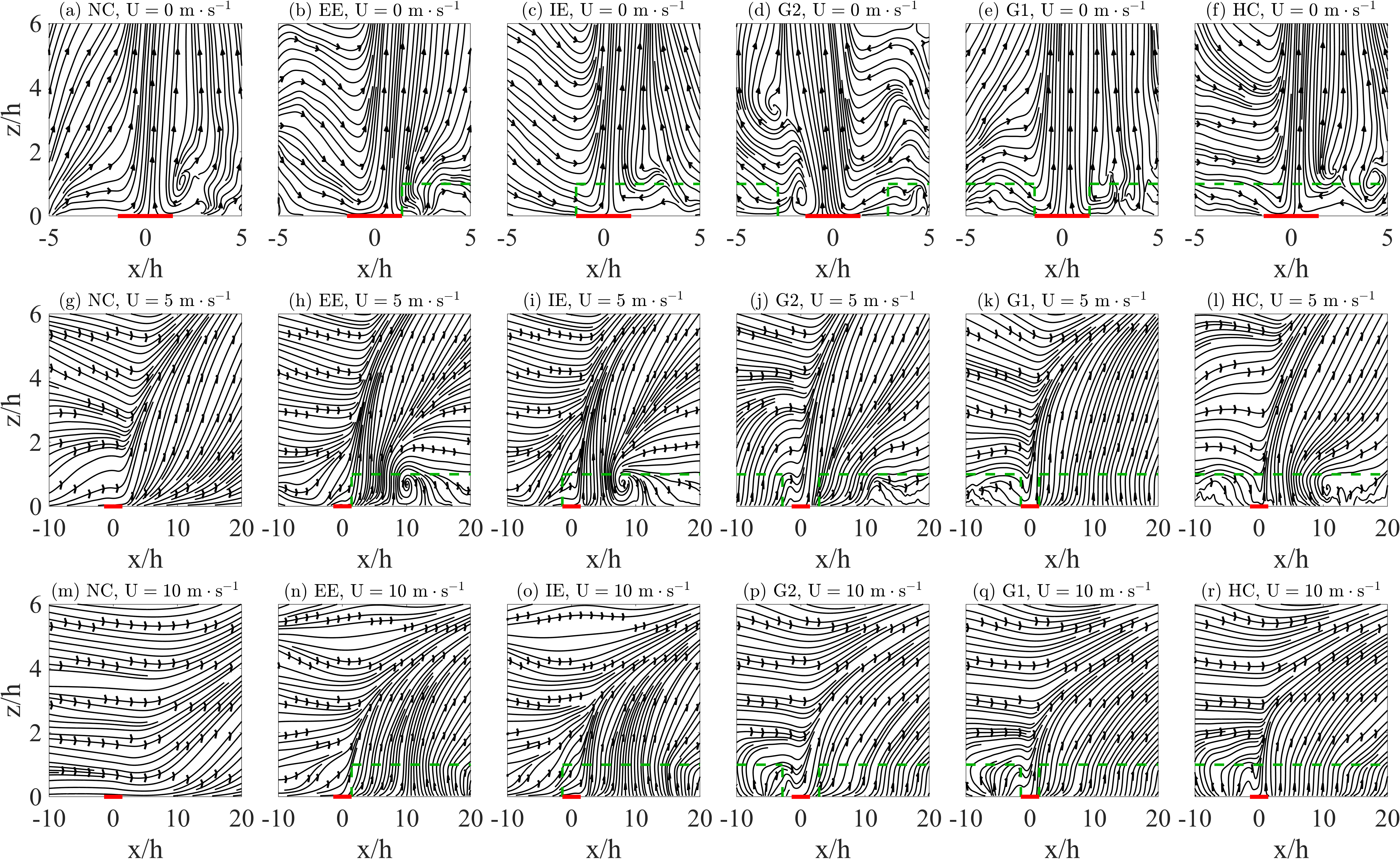}
    \end{subfigure}
    \caption{Velocity streamlines in the XZ-plane at $y = 500 \ \mathrm{m}$ for the wind speed cases of $\mathrm{U = 0 \ m \cdot s^{-1}}$ (top row), $\mathrm{U = 5 \ m \cdot s^{-1}}$ (middle row), and $\mathrm{U = 10 \ m \cdot s^{-1}}$ (bottom row) for each canopy configuration with plume. The x-axis origin is located at the center of the heat patch, and both the x and z axes are normalized by the canopy height ($h$).}
    \label{fig:All plume Streamlines}
\end{figure}

% Observed plume behavior and possible influence of top boundary 
The no wind case exhibited strong vertical updrafts driven by the high surface heat flux. As a response, entrainment of air toward the center of the patch created crosswinds that interacted with the canopy depending on the configuration. As shown in Fig. \ref{fig:All plume Streamlines}, the streamlines for the no wind case show variation of flow on either side of the heat patch, indicated as a red line on the bottom of each panel. For example, Fig. \ref{fig:All plume Streamlines} for NC, G2, and HC in the top row displayed a recirculation only on one side of the heat patch. It should be noted that the plume interacts with the top boundary of the domain. In PALM, vertical velocity $w$ is assumed to be zero at the top wall \citep{maronga_parallelized_2015}, imposing Neumann boundary conditions that modify pressure at that surface. This effectively creates a barrier that the plume impinges on, creating large scale structures on either side. These structures interact with the plume and each other due to the cyclic boundary conditions applied at the inlet and outlet for these cases. These exchanges can be seen in Fig. \ref{fig:U0 streamlines zoomed out} in the appendix. However, the velocity of these structures were much lower than the plume's horizontal and vertical velocity, as shown in Fig. \ref{fig:U0 spatial mean variables} in the appendix.

\begin{figure}[htbp]
    \centering
    \begin{subfigure}{0.75\textwidth}
        % \caption{}
        \includegraphics[width=\linewidth]{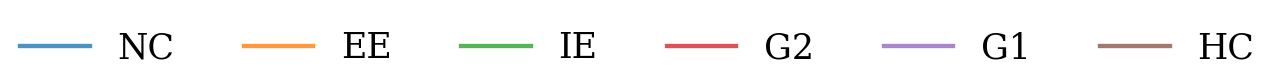}
        % \label{fig:U0 u vertical z/h=-3}
    \end{subfigure}
    \hfill
    \begin{subfigure}{\textwidth}
        % \caption{}
        \includegraphics[width=\linewidth]{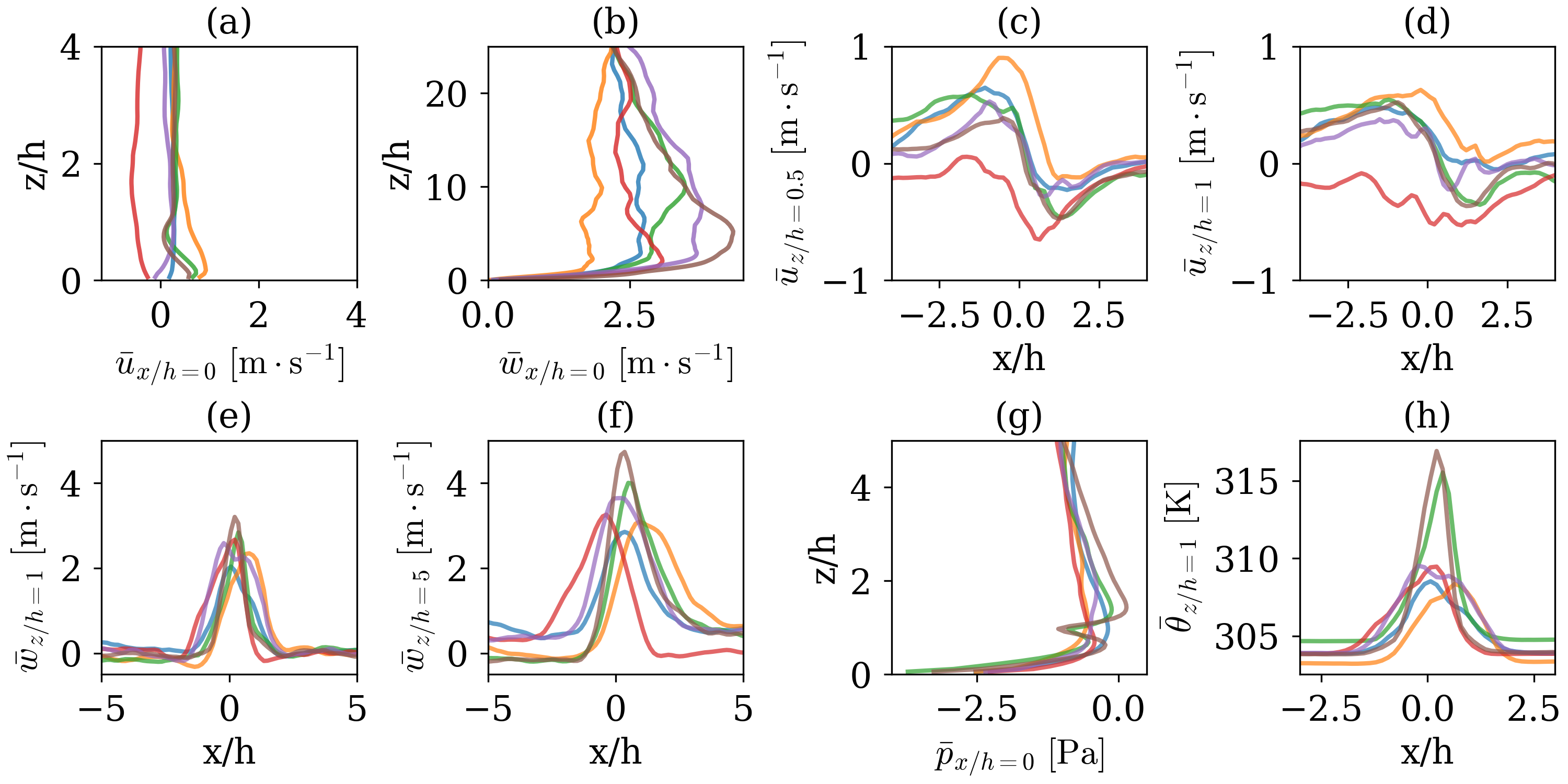}
        % \label{fig:U0 u vertical z/h=0}
    \end{subfigure}
    \caption{Profiles of mean flow variables for the no-wind case $(\mathrm{U=0 \ m \cdot s^{-1}})$. Vertical profiles of $\bar{u}$ (a) and $\bar{w}$ (b) at $x/h=0$; horizontal profiles of $\bar{u}$ at $z/h=0.5$ (c) and $z/h=1$ (d); horizontal profiles of $\bar{w}$ at $z/h=1$ (e) and $z/h=5$ (f); vertical profiles of $\bar{p}$ (g); and horizontal profiles of $\bar{\theta}$ at $z/h=1$ (h).}
    \label{fig:U0 mean variables}
\end{figure}

The flow statistics associated with plume development under no crosswind are presented in Fig. \ref{fig:U0 mean variables}. Non-zero horizontal velocity is observed below the canopy top where flow is entrained by the updrafts of the plume (Fig. \ref{fig:U0 mean variables}a). In the EE configuration, the positive $u$ velocity on the clearing side (left) is greater than the negative $u$ velocity on the canopy side (right), as shown in Fig. \ref{fig:U0 mean variables}c, d. The remaining canopies appear to have similar positive and negative $u$ velocities on the left and right sides of the plume (Fig. \ref{fig:U0 spatial mean variables}).

% Streamlines

% Vertical velocity differences
The largest vertical velocity (Figure \ref{fig:U0 mean variables}b) is observed in the homogeneous canopy (HC) configuration. The vertical velocity at the plume center rapidly accelerates below the canopy top, then peaks around 5-10h height, and gradually decreases with height above. In some cases, such as EE and NC, the decrease with height is much slower, with the EE canopy configuration showing a slight increase. Each canopy configuration varied in peak vertical velocity as shown in Fig. \ref{fig:U0 mean variables}b, e, and f. The homogeneous canopy produced the largest, followed by the IE configuration, while the NC and EE configurations showed the lowest. In the HC and IE canopies, Fig. \ref{fig:U0 mean variables}g shows an abrupt change in the vertical pressure profile. At the surface, low pressure is generated by the buoyancy source, which then increases through the canopy with height. At the canopy top, there is a rapid pressure drop followed by a rapid increase above. This abrupt transition may contribute to higher vertical velocities. Temperature is highest in the HC and IE configurations (Fig. \ref{fig:U0 mean variables}h and Fig. \ref{fig:U0 spatial mean variables} in the appendix), and shows a narrower distribution compared to the other cases.

\begin{figure}[htbp]
    \centering
    \begin{subfigure}{0.75\textwidth}
        % \caption{}
        \includegraphics[width=\linewidth]{Figures/all_cases/legend.png}
        % \label{fig:U10 u vertical z/h=-3}
    \end{subfigure}
    \hfill
    \begin{subfigure}{\textwidth}
        % \caption{}
        \includegraphics[width=\linewidth]{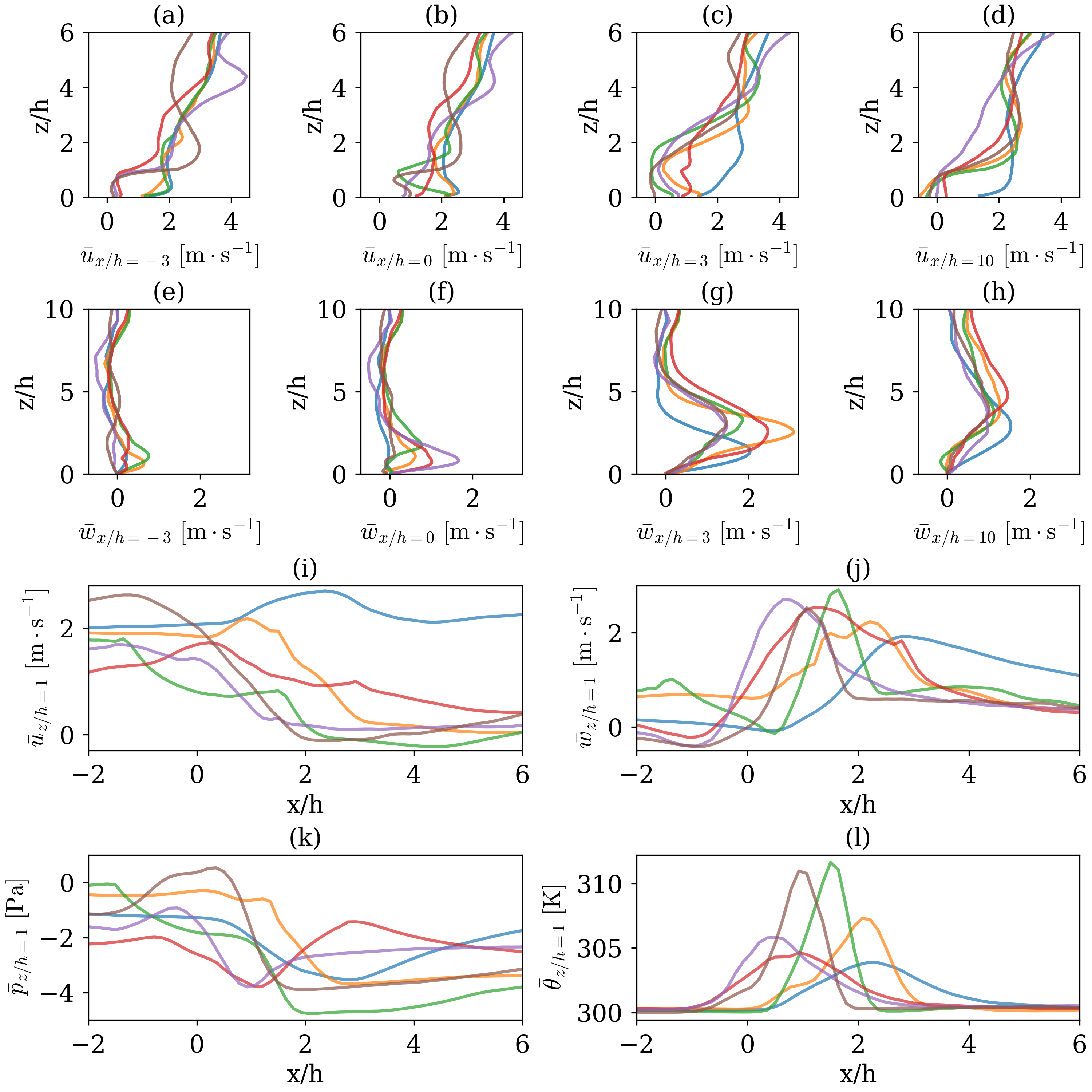}
        % \label{fig:U10 u vertical z/h=0}
    \end{subfigure}
    \caption{Profiles of mean flow variables for the crosswind case $(\mathrm{U=5 \ m \cdot s^{-1}})$. Vertical profiles of $\bar{u}$ (a-d) and $\bar{w}$ (e-h) at $x/h=-3, \ 0, \ 3, \ \text{and} \ 10$; horizontal profiles of $\bar{u}$ (i) and $\bar{w}$ at $z/h=1$ (j); horizontal profiles of $\bar{p}$ (k) and $\bar{\theta}$ at $z/h=1$ (l).}
    \label{fig:U5 mean variables}
\end{figure}

\subsubsection{Plume in Crosswind $(5 \ \mathrm{m \cdot s^{-1}})$}

Additional complexities beyond the baseline canopy flow evidently characterize the presence of a plume in crosswind. Plume induced modulations of the recirculation zones, plume tilt, and bulk flow structures can be visually delineated from Fig. \ref{fig:All plume Streamlines}. The first-order consequence of a buoyant plume is the generation of strong updrafts, which lead to entrainment flows that are, in turn, impacted by the canopy structure. The recirculation zones previously seen in the baseline flows were effectively destroyed by the plume updrafts. New recirculation zones formed in the EE, IE, and HC cases near $x/h = 10$. There was a noticeable shift in the recirculation zone between the EE and IE configurations, where the EE recirculation is shifted by about 2h downstream from the IE case. This may be because, in the EE case, the upstream air mass is first lifted by the plume before interacting with the edge. Whereas, in the IE case, the upstream air parcel is lifted due to edge effects before being influenced by the plume.

% Plume tilt
The difference in plume tilt between all the canopy configurations is seen in Fig. \ref{fig:U5 mean variables}j which shows the horizontal profiles of mean vertical velocity $\bar{w}$ distributions at the height of the canopy. The plume tilt can be inferred from tracking the mean vertical velocity, since a steeper and more vertical plume would be characterized by a higher $\bar{w}$ \citep{devenish_entrainment_2010}. Using the horizontal profile of $\bar{w}$, the plume in the IE configuration is observed to have a steeper tilt than the EE plume at the canopy top, although it is not very clear from the visualizations of the streamlines in Fig. \ref{fig:All plume Streamlines}.  The canopy configuration with the smaller (G1:100m) gap is associated with a slightly steeper tilt, while the plume for the no canopy case shows the strongest tilt, which is evidently due to the lack of canopy induced drag.

\begin{figure}[htbp]
    \centering
    \begin{subfigure}{0.65\textwidth}
        \caption{Isometric View}
        \includegraphics[width=\linewidth]{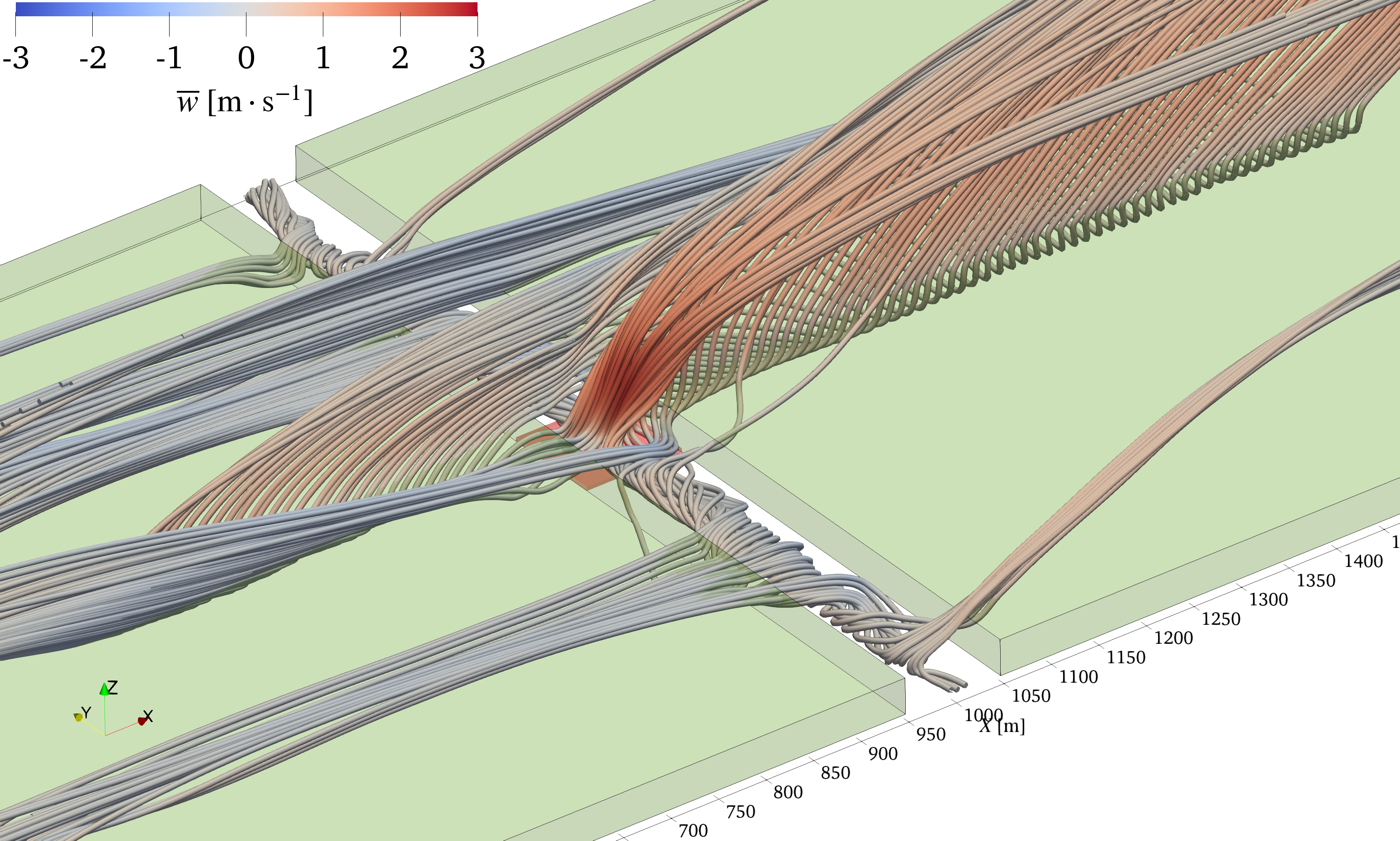}
        \label{fig:U5 Paraview iso}
    \end{subfigure}
    \hfill
    \begin{subfigure}{0.65\textwidth}
        \caption{Transverse View}
        \includegraphics[width=\linewidth]{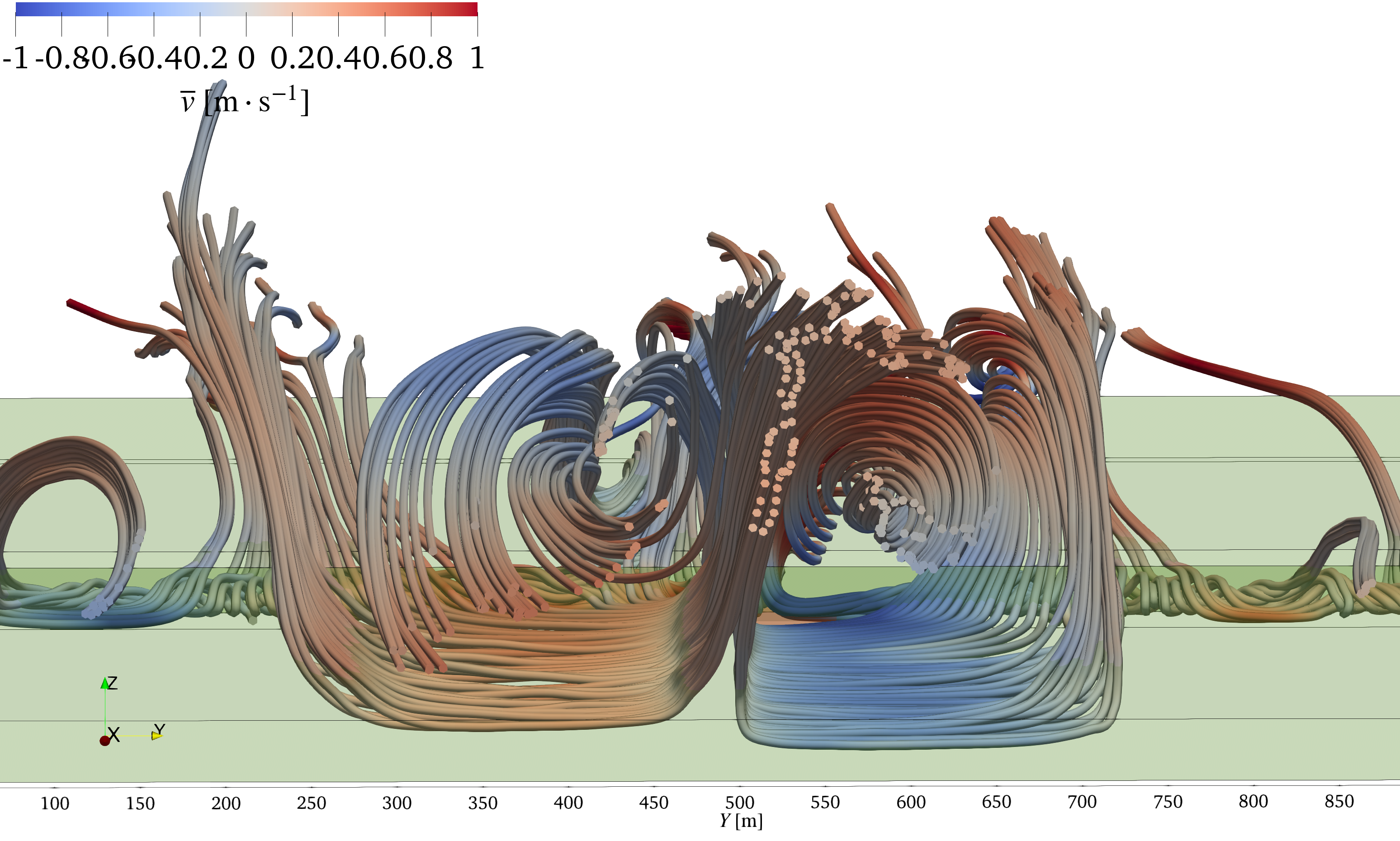}
        \label{fig:U5 Paraview transverse}
    \end{subfigure}
    \hfill
    \begin{subfigure}{0.65\textwidth}
        \caption{Q-Criterion}
        \includegraphics[width=\linewidth]{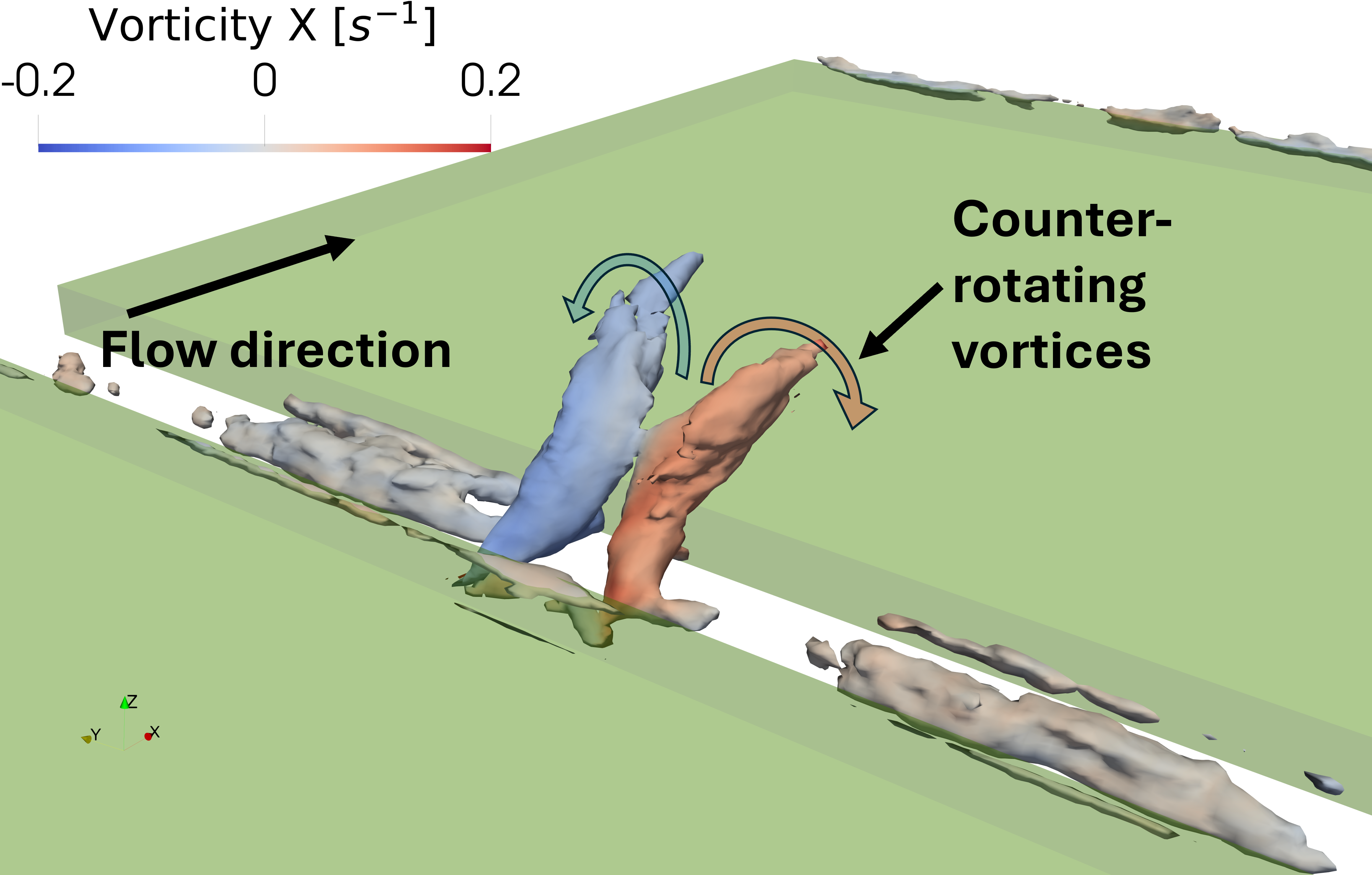}
        \label{fig:U5 Q-criterion}
    \end{subfigure}
    \caption{Three-dimensional visualization of the G1 crosswind case $(\mathrm{U=5 \ m \cdot s^{-1}})$ with plume; streamlines colored by $\bar{w}$ (a) , $\bar{v}$  (b), and Q-criterion iso-surface (c) of Q=0.001 colored by vorticity in x-direction. Plots are visualized in ParaView, with green-shaded blocks representing the canopy layout.}
    \label{fig:U5 Paraview}
\end{figure}

% velocity reduction
As the crosswind interacts with the plume and the canopy, the streamwise velocity is reduced further downstream. The vertical profiles of mean horizontal velocity $\bar{u}$ in Figs. \ref{fig:U5 mean variables}b and c show the reduction as wind flows past the heat patch (located at $x/h=0$). The profile's inflection point is lifted vertically, indicating a reduction of wind speed by the plume. Similar observations are seen in Fig. \ref{fig:U5 mean variables}i, where the horizontal profile of $\bar{u}$ at the canopy top is decreased after the heat patch. On the other hand, the vertical velocity increases in speed from the upstream vertical profile at $x/h=-3$ to the downstream profile at $x/h=10$ by the plume (Figs. \ref{fig:U5 mean variables}e-h).

% Pressure and velocity
The horizontal profiles of mean pressure $\bar{p}$ across the canopy top are shown in Fig. \ref{fig:U5 mean variables}k. The impact of the plume is analogous to a backward facing step (although porous), since the pressure is found to increase at the windward side of the plume and drop at the leeward side of the plume (which leads to plume induced entrainment). The exit edge of a gap is also characterized with a drop of pressure and therefore, the G2 case is characterized by a milder drop in pressure due to the opposing impacts of the exit edge and the plume induced pressure rise before and at $x/h=0$. On the other hand, the G2 case experiences the entrance edge effects immediately downstream of the plume, and therefore characterized by a sharp rise of pressure downstream of the patch center. The G1 configuration, although associated with a canopy gap, is mainly dominated by the buoyant plume dynamics (since the heat patch occupies the entire gap), and the flow does not have sufficient room to develop downstream to experience the impact of the entrance edge of the canopy. For the other cases, the pressure drops at the leeward side of the plume and continues to rise slowly downstream. The largest change in pressure is seen in the IE case. Fig. \ref{fig:U5 spatial mean variables} in the appendix shows the spatial variation of pressure gradient location across the canopy configurations. 

% Temperature
Temperature at the canopy top varies in distribution, as shown in Fig. \ref{fig:U5 mean variables}l. Similar to the no wind cases, the highest temperatures were observed in IE and HC configurations. The location of peak temperature distribution is similar to the peak of the vertical velocity distributions in Fig. \ref{fig:U5 mean variables}j. As such, the canopy configurations, IE and HC, seem to have the largest temperatures at the canopy top. These differences between the canopy configurations are visualized spatially in Fig. \ref{fig:U5 spatial mean variables} in the appendix, which shows high temperatures at the canopy top in the IE and HC configurations.

% Complex structures
The plume-canopy interactions are visualized in three dimensions for the G1 configuration in Fig. \ref{fig:U5 Paraview}. In the isometric view of Fig. \ref{fig:U5 Paraview iso}, the recirculation zones along the lateral ends of the heat patch are seen to be pulled in towards the plume. Above the heat patch, high vertical velocity is seen to drive the flow upward. As a result of this, the upstream flow collides with the plume and twists around it. Figure \ref{fig:U5 Paraview transverse} shows a transverse view of the streamlines colored by the lateral velocity $v$. It can be seen that counter-rotating vortices are formed in the downstream of the plume. Figure \ref{fig:U5 Q-criterion} shows the q-criterion iso-surface of Q=0.001 colored by vorticity in the x-direction. It can be seen that the counter-rotating vortices formed at the base of the plume.

\subsubsection{Plume in Crosswind $(10 \ \mathrm{m \cdot s^{-1}})$}
% Intro
At 10 $\mathrm{m \cdot s^{-1}}$ wind speed, the most notable changes from the previous wind speed are that the recirculation zones are no longer present, there is less horizontal velocity reduction, the plume is more tilted, and pressure gradients for the edge configuration are larger.

% Recirculations, velocity reduction
The recirculation zones previously seen in the EE, IE, and HC canopies at 5 $\mathrm{m \cdot s^{-1}}$ wind speed were no longer observed at 10 $\mathrm{m \cdot s^{-1}}$ (see Fig. \ref{fig:All plume Streamlines}). Instead, there were persistent updrafts along the downstream side of the heat patch. As seen in Fig.  \ref{fig:U10 mean variables}i-j in the appendix, there is less streamwise velocity reduction at the canopy top across the heat patch. The no-canopy configuration appeared to have stronger streamwise wind velocity compared to its plume updraft velocity. 

% Plume tilt
The streamlines in Fig. \ref{fig:All plume Streamlines} show that the NC plume updrafts were more tilted than the previous wind speed. The same is true for the rest of the configurations, as shown in Fig. \ref{fig:U10 mean variables}j in the appendix; the peak vertical velocity is shifted further downstream than the 5 $\mathrm{m \cdot s^{-1}}$ wind speed, indicating that the plumes are more tilted. The vertical profiles of $\bar{u}$ in Figs. \ref{fig:U10 mean variables}a–d in the appendix show minimal variability above the canopy (above $z/h=2$) across all canopy configurations, except for the NC case as noted earlier. Further downstream at $x/h=10$, differences occur between the EE/IE cases compared to the G1/G2/HC cases. For $\bar{w}$ (Figs. \ref{fig:U10 mean variables}e–h in the appendix), the peaks of the distributions are found between 2–4h in height. This extended plume tilt is illustrated in Figs. \ref{fig:U10 spatial mean variables} in the appendix.

% Pressure, Temperature
The pressure gradients for EE/IE configurations were observed to be considerably larger than the rest of the configurations (Fig. \ref{fig:U10 mean variables}k in the appendix). Additionally, the temperature at the canopy top is significantly higher in the HC case compared to the other canopy configurations (Fig. \ref{fig:U10 mean variables}l in the appendix).

\subsection{Turbulence Statistics}

\subsubsection{Plume in No Crosswind $(0 \ \mathrm{m \cdot s^{-1}})$}
\begin{figure}[htbp]
    \centering
    % Legend on its own line
    \includegraphics[width=0.75\textwidth]{Figures/all_cases/legend.png}
    
    % \vspace{0.5em}

    % --- Top row: Momentum flux ---
    \begin{subfigure}{0.3\textwidth}
        \caption{$0 \ \mathrm{m \cdot s^{-1}}$}
        \includegraphics[trim={0 0 0 1.5em},clip,width=\linewidth]{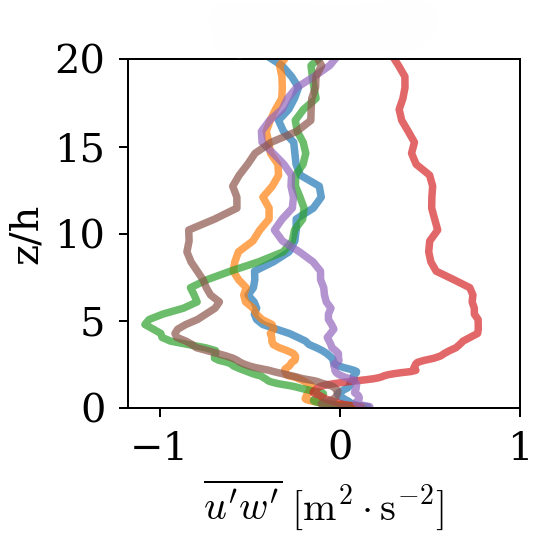}
    \end{subfigure}
    \hfill
    \begin{subfigure}{0.3\textwidth}
        \caption{$5 \ \mathrm{m \cdot s^{-1}}$}
        \includegraphics[trim={0 0 0 1.5em},clip,width=\linewidth]{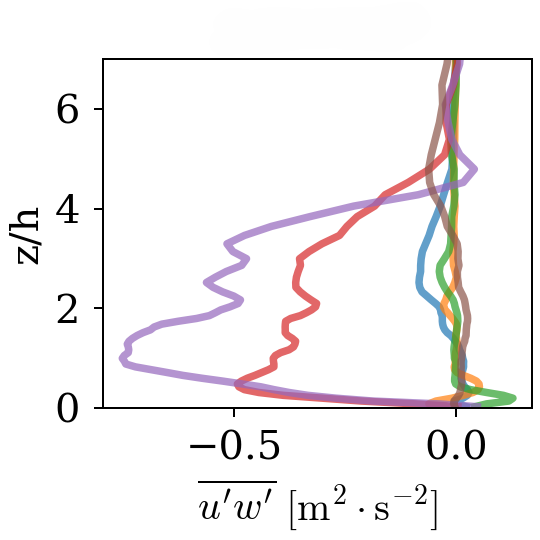}
    \end{subfigure}
    \hfill
    \begin{subfigure}{0.3\textwidth}
        \caption{$10 \ \mathrm{m \cdot s^{-1}}$}
        \includegraphics[trim={0 0 0 1.5em},clip,width=\linewidth]{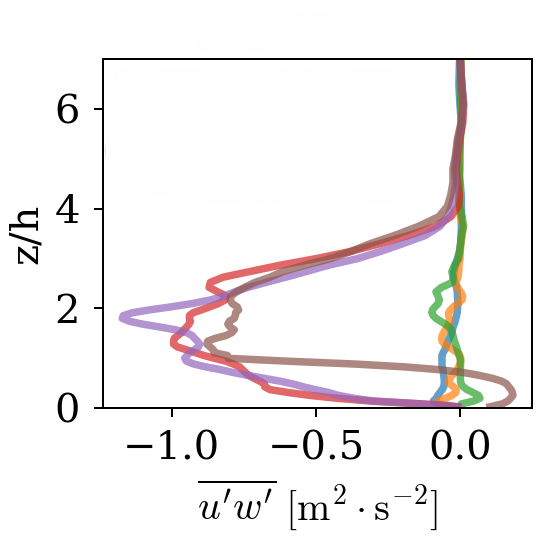}
    \end{subfigure}

    \vfill

    % --- Bottom row: TKE ---
    \begin{subfigure}{0.3\textwidth}
        \caption{$0 \ \mathrm{m \cdot s^{-1}}$}
        \includegraphics[width=\linewidth]{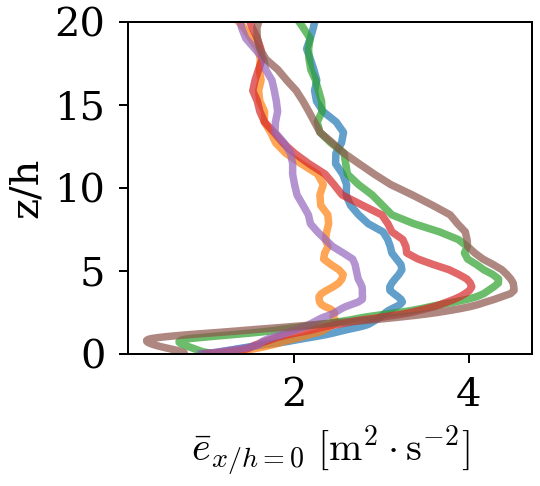}
    \end{subfigure}
    \hfill
    \begin{subfigure}{0.3\textwidth}
        \caption{$5 \ \mathrm{m \cdot s^{-1}}$}
        \includegraphics[width=\linewidth]{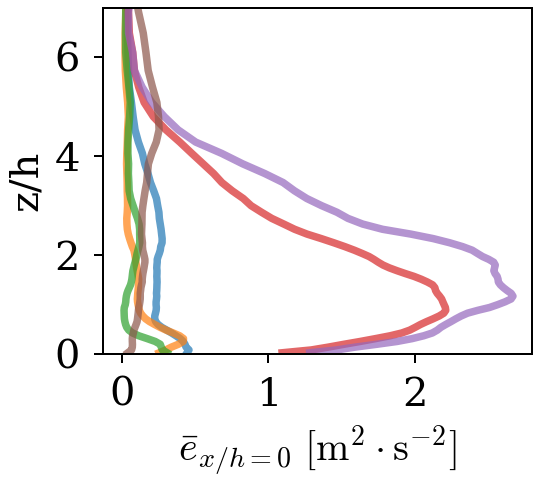}
    \end{subfigure}
    \hfill
    \begin{subfigure}{0.3\textwidth}
        \caption{$10 \ \mathrm{m \cdot s^{-1}}$}
        \includegraphics[width=\linewidth]{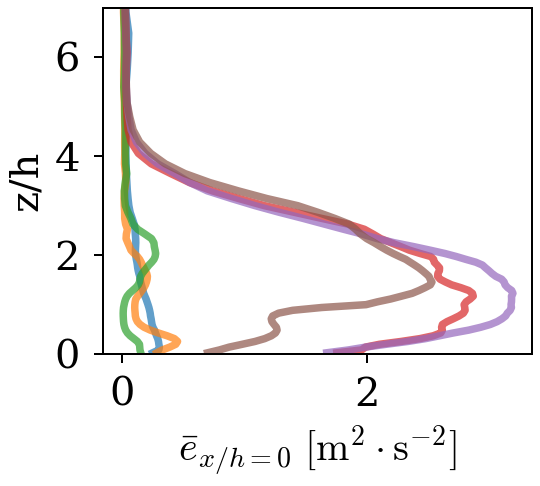}
    \end{subfigure}

    \caption{Profiles of mean momentum flux (top row) and turbulent kinetic energy (bottom) for the no-wind case $(\mathrm{U=0 \ m \cdot s^{-1}})$, and crosswind cases $(\mathrm{U=5 \ m \cdot s^{-1}})$ and $(\mathrm{U=10 \ m \cdot s^{-1}})$.}
    \label{fig:All momentum flux and TKE profiles}
\end{figure}

\begin{figure}[ht!]
    \centering
    \begin{subfigure}{0.25\textwidth}
        % \caption{}
        \includegraphics[width=\linewidth]{Figures/all_cases/spatial_plot_legend.png}
    \end{subfigure}
    
    \vspace{0.5em}
    
    \begin{subfigure}{\textwidth}
        % \caption{}
        \includegraphics[width=\linewidth]{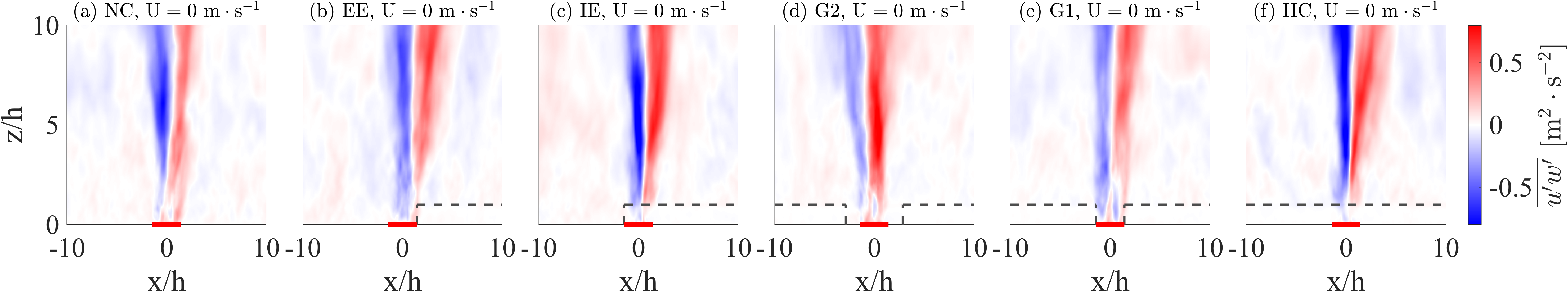}
    \end{subfigure}

    \vspace{0.5em}

    \begin{subfigure}{\textwidth}
        % \caption{}
        \includegraphics[width=\linewidth]{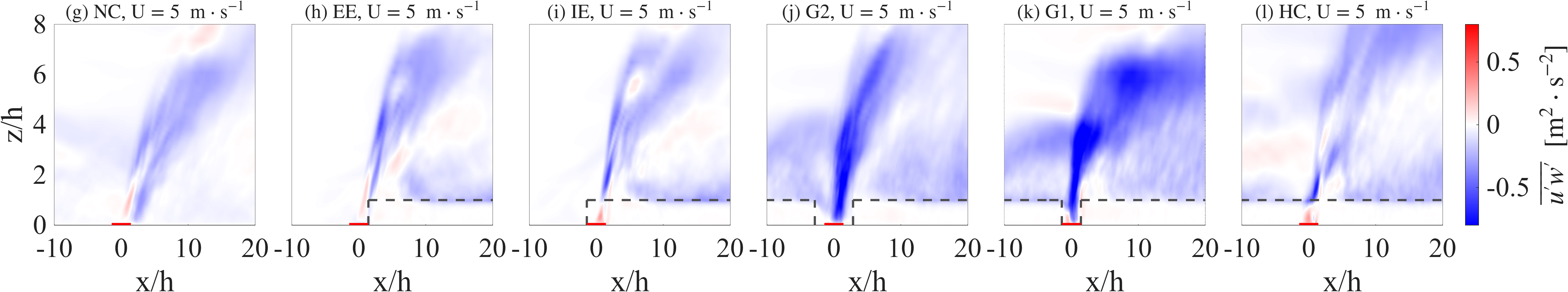}
    \end{subfigure}

    \vspace{0.5em}

    \begin{subfigure}{\textwidth}
        % \caption{}
        \includegraphics[width=\linewidth]{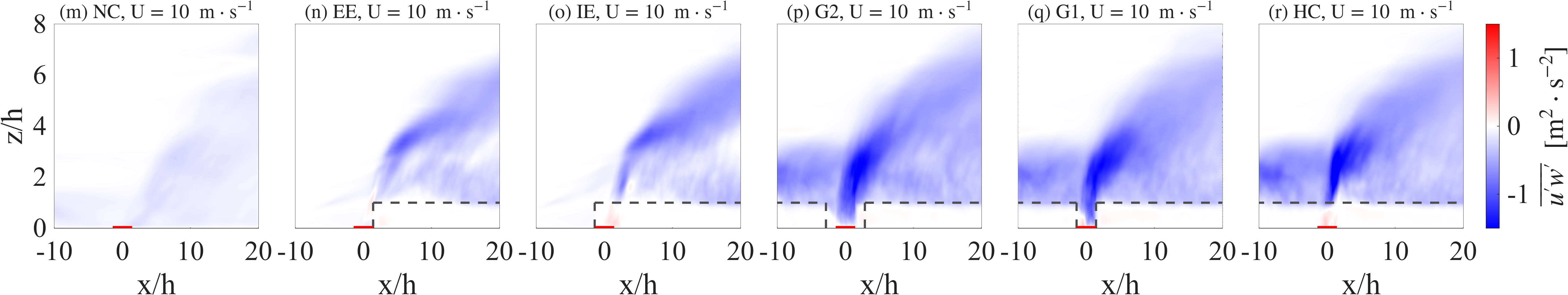}
    \end{subfigure}
    \caption{Spatial distribution of momentum flux in the XZ-plane at $y = 500 \ \mathrm{m}$ for the wind speed cases of $\mathrm{U = 0 \ m \cdot s^{-1}}$ (top row), $\mathrm{U = 5 \ m \cdot s^{-1}}$ (middle row), and $\mathrm{U = 10 \ m \cdot s^{-1}}$ (bottom row) for each canopy configuration with plume. The x-axis origin is located at the center of the heat patch, and both the x and z axes are normalized by the canopy height ($h$).}
    \label{fig:All momentum flux spatial}
\end{figure}

The analysis of turbulence behavior for the no-wind case begins with the momentum flux. Vertical profiles above the center of the heat patch, in Fig. \ref{fig:All momentum flux and TKE profiles}a, indicate that momentum is reduced with height to about 4-6h height, except for the G2 configuration, which shows an increase, before settling to a steady state further aloft. However, the spatial momentum flux plots in Fig. \ref{fig:All momentum flux spatial}a-f illustrate a bipolar negative and positive momentum flux on the left and right sides, respectively, of the plume. 

\begin{figure}[ht!]
    \centering
    \begin{subfigure}{0.25\textwidth}
        % \caption{}
        \includegraphics[width=\linewidth]{Figures/all_cases/spatial_plot_legend.png}
    \end{subfigure}
    \hfill
    \begin{subfigure}{\textwidth}
        % \caption{}
        \includegraphics[width=\linewidth]{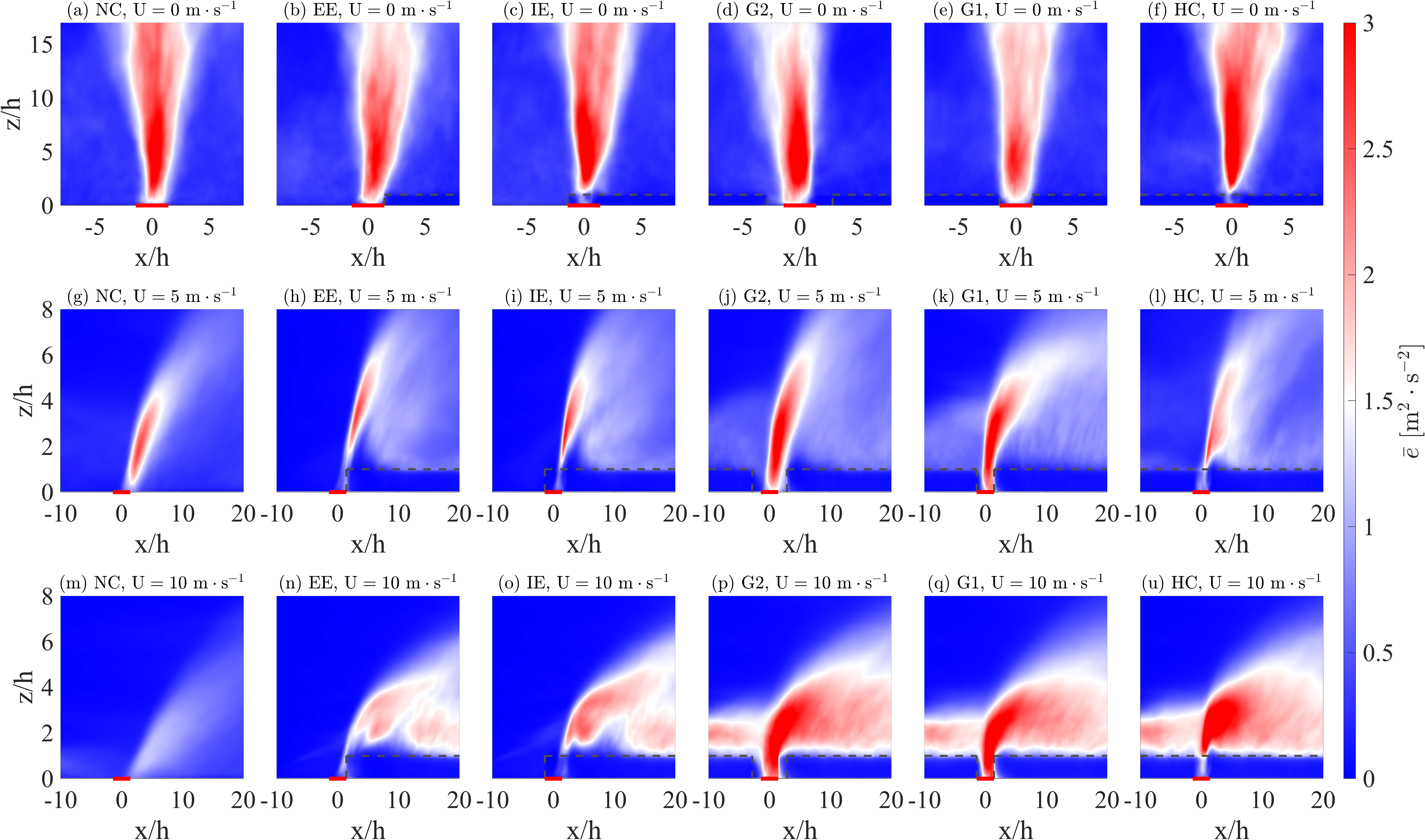}
    \end{subfigure}
    \caption{Spatial distribution of turbulent kinetic energy in the XZ-plane at $y = 500 \ \mathrm{m}$ for the wind speed cases of $\mathrm{U = 0 \ m \cdot s^{-1}}$ (top row), $\mathrm{U = 5 \ m \cdot s^{-1}}$ (middle row), and $\mathrm{U = 10 \ m \cdot s^{-1}}$ (bottom row) for each canopy configuration with plume. The x-axis origin is located at the center of the heat patch, and both the x and z axes are normalized by the canopy height ($h$).}
    \label{fig:All TKE spatial}
\end{figure}

Vertical profiles of turbulent kinetic energy taken at the center of the plume were also found to increase with height to a maximum at 4-6h height, as shown in Fig. \ref{fig:All momentum flux and TKE profiles}d. The homogeneous and internal edge canopy configurations show a reduction in TKE at the canopy top. Above the canopy, however, both configurations exhibit a rapid increase, reaching the largest TKE values among the six canopy configurations. As TKE reaches its maximum above the canopy, its gradually reduces with height. Figure \ref{fig:All TKE spatial}a-f exhibits this reduction in TKE with height and the spatial variation in TKE among the canopy configurations. 

\begin{figure}[ht!]
    \centering

    \begin{subfigure}{0.65\textwidth}
        \includegraphics[width=\linewidth]{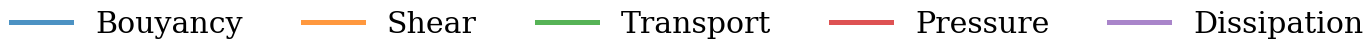}
    \end{subfigure}
    
    \begin{subfigure}{\textwidth}
    \includegraphics[width=\textwidth]{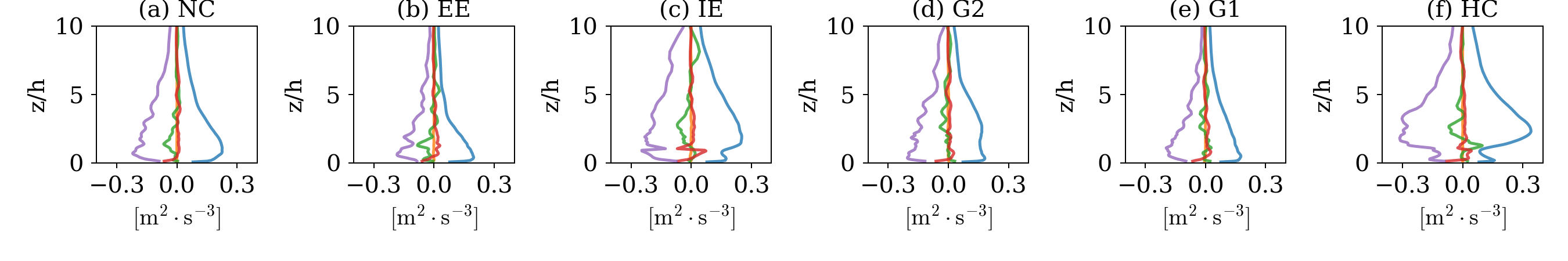}
    \end{subfigure}

    \begin{subfigure}{\textwidth}
    \includegraphics[width=\textwidth]{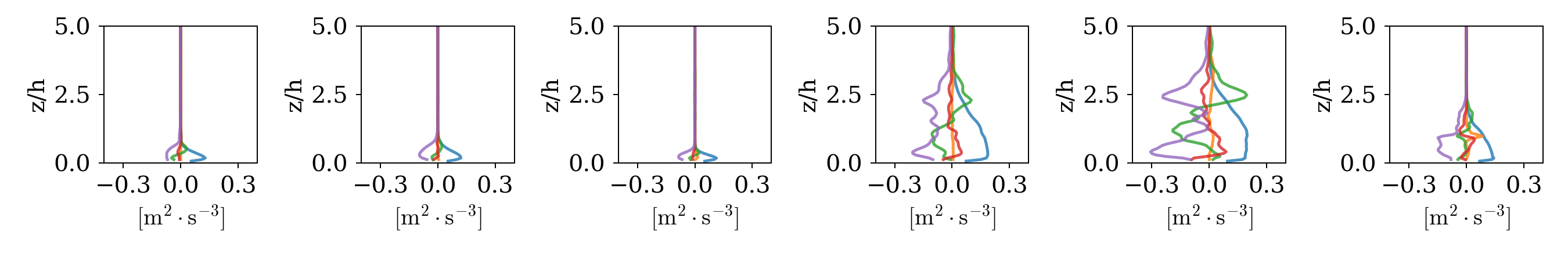}
    \end{subfigure}

    \begin{subfigure}{\textwidth}
    \includegraphics[width=\textwidth]{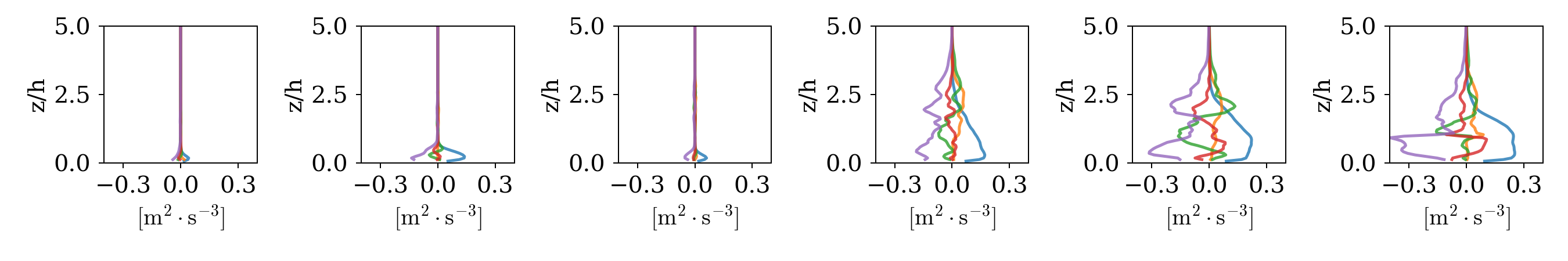}
    \end{subfigure}
    \caption{Vertical profiles of TKE budget terms at the center of the heat patch for the wind speed cases of $\mathrm{U = 0 \ m \cdot s^{-1}}$ (top row), $\mathrm{U = 5 \ m \cdot s^{-1}}$ (middle row), and $\mathrm{U = 10 \ m \cdot s^{-1}}$ (bottom row) for each canopy configuration with plume.}
    \label{fig:All TKE Budget Profiles}
\end{figure}

Vertical profiles of the TKE budget terms, also taken at the plume center, are shown in Fig. \ref{fig:All TKE Budget Profiles}a-f. Buoyancy was found to be the dominant source of TKE production as there was no imposed crosswind to interact with the plume. Buoyant production is noticed to be similar in the NC, EE, G2, and G1 cases. These cases feature no canopy directly above the heat patch, and thus, the plume freely develops vertically upwards. In the IE and HC cases, buoyant production profiles show a reduction at the canopy top, then enhances at 2h. Shear production, turbulent transport, and pressure correlation are similar across all cases. There is a noticeable trend of downward TKE transport above 2h, and upward TKE transport below that height. Pressure correlation is fairly similar across all the cases. The IE and HC cases exhibit positive values below the canopy, then transition quickly to negative values at the canopy top. Dissipation was computed from the summation of remaining budget terms, with the assumption of no residuals.

%% Momentum flux spatial plots

% TKE spatial plots

% Budget Term spatial plots for U10
\begin{figure}[ht!]
    \centering
    \begin{subfigure}{0.25\textwidth}
        % \caption{}
        \includegraphics[width=\linewidth]{Figures/all_cases/spatial_plot_legend.png}
    \end{subfigure}
    
    \vspace{0.5em}
    
    \begin{subfigure}{\textwidth}
        % \caption{}
        \includegraphics[width=\linewidth]{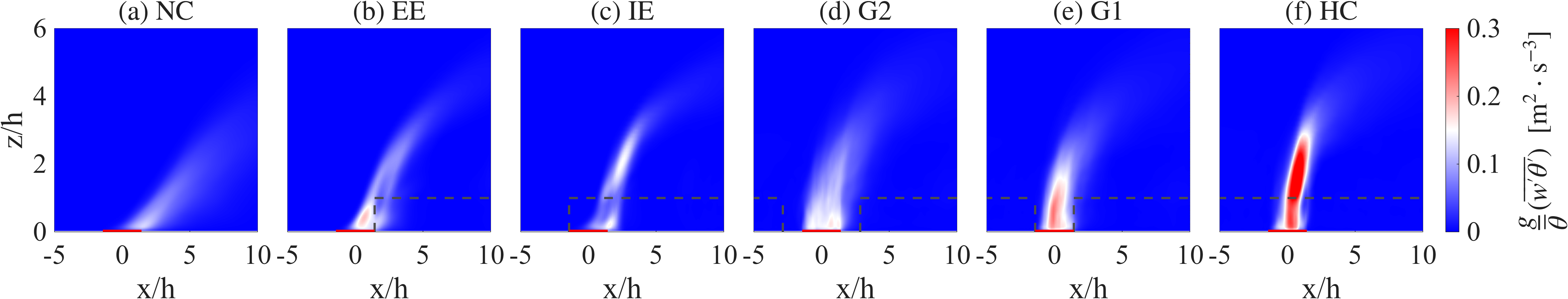}
    \end{subfigure}

    \vspace{0.5em}
    
    \begin{subfigure}{\textwidth}
        % \caption{}
        \includegraphics[width=\linewidth]{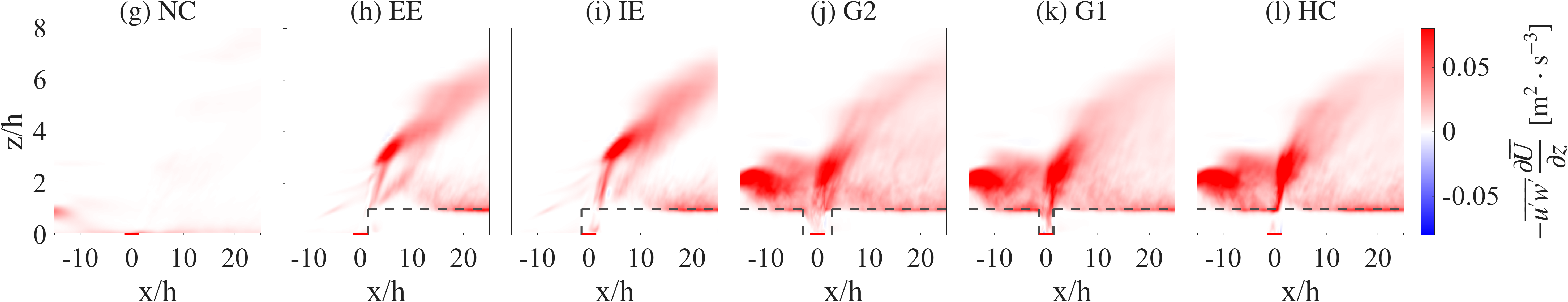}
    \end{subfigure}

    \vspace{0.5em}
    
    \begin{subfigure}{\textwidth}
        % \caption{}
        \includegraphics[width=\linewidth]{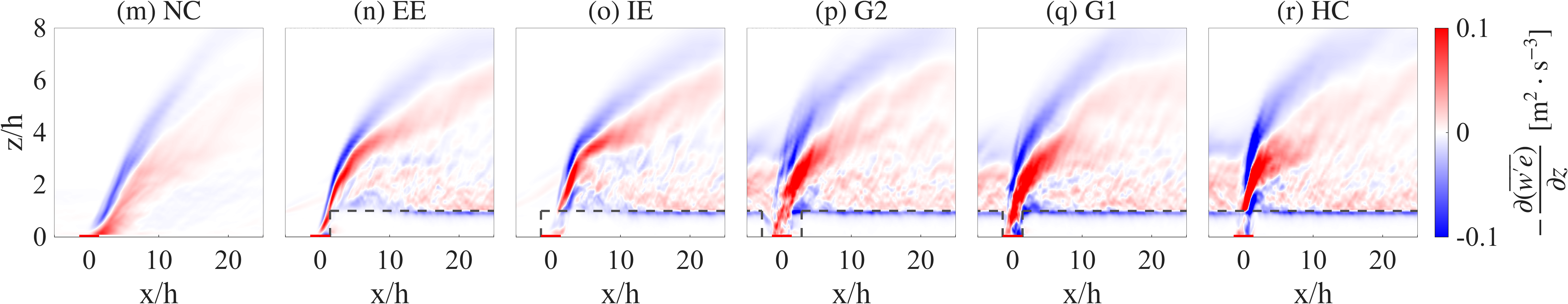}
    \end{subfigure}
    \caption{Spatial distribution of TKE budget terms in the XZ-plane at $y = 500 \ \mathrm{m}$ for the $\mathrm{U = 10 \ m \cdot s^{-1}}$ wind speed case of buoyancy (top row), shear (middle row), and transport (bottom row) for each canopy configuration with plume. The x-axis origin is located at the center of the heat patch, and both the x and z axes are normalized by the canopy height ($h$).}
    \label{fig:U10 TKE Budget Spatial}
\end{figure}

\subsubsection{Plume in Crosswind $(5 \ \mathrm{m \cdot s^{-1}})$}

Plumes in 5 $\mathrm{m \cdot s^{-1}}$ crosswinds exhibit downward momentum flux above the canopy, and on the windward side of the plume. The leeward side of the plume is associated with upward vertical momentum flux. Notice that unlike the non-plume situation, the presence of the heat source below the canopy injects momentum into the flow from below, which is then modulated depending on the flow configuration. Figure \ref{fig:All momentum flux spatial} displays the spatial variation of momentum reduction by the plume across each canopy configuration. The gap canopy configurations were observed to produce the strongest downward momentum transport, as seen in Fig. \ref{fig:All momentum flux and TKE profiles}b.

The highest TKE is observed in the 100 m and 200 m gap canopy cases. As shown in Figure \ref{fig:All momentum flux and TKE profiles}e, TKE peaks at 1-2 canopy heights $h$ above the heat patch. Spatial plots of TKE in the XZ plane (Figure \ref{fig:All TKE spatial}d-e) indicate that the gap canopies vary from the rest of the cases. The NC, EE, IE, and HC cases show comparable regions of TKE above the canopy layer, whereas the gap cases exhibit TKE under the canopy layer and within the bounds of the gaps. 

Examining the regions of the gaps, it can be observed that there is more TKE closer to the heat patch. TKE budget terms show enhanced buoyant production in the gap cases (Figure \ref{fig:All TKE Budget Profiles}j-k) in contrast to the rest of the canopy configurations. Buoyant production levels in the G1 case are maintained from the surface to near the canopy top, whereas the G2 case shows a gradual reduction with height. The budgets in the rest of the cases show low TKE production at the center of the heat patch. This is likely due to the crosswind pushing the plume downstream, resulting in less TKE above the heat patch. Shear production is dominated by buoyancy across all cases. However, the stronger shear production, as expected, is observed close to the canopy top, especially for the HC case. Downward TKE transport is observed below the canopy and upwards above the canopy. This effect is enhanced in the gap canopy cases. Pressure correlation is positive below the canopy and negative above. Dissipation follows a similar trend as the no wind cases in that it mirrors the buoyancy; however, there are more differences observed in the G2, G1, and HC cases.

\subsubsection{Plume in Crosswind $(10 \ \mathrm{m \cdot s^{-1}})$}

Higher crosswind speeds resulted in enhanced downward momentum transport at the windward side of the plume as seen in Fig. \ref{fig:All momentum flux and TKE profiles}c. The homogeneous canopy configuration is associated with similar levels of downward momentum flux as the gap canopy configurations. The impact of momentum injection from below is less pronounced because of stronger downward momentum transport, as evident by the smaller extent of positive momentum flux regions inside and adjacent to the canopy, as well as the downside edge of the plume.

Figure \ref{fig:All momentum flux and TKE profiles}f shows larger peak TKE values above the center of the patch than the 5 $\mathrm{m \cdot s^{-1}}$ case. There is a noticeable increase in peak TKE in the HC configuration. The spatial distributions of TKE in Fig. \ref{fig:All TKE spatial}m, show that the no canopy configuration reduced TKE at the higher wind speeds. TKE is also reduced and more dispersed downstream in the edge cases (EE/IE). The gap canopies (G2/G1) and homogeneous canopies show an increase in TKE both in the upstream/downstream locations and within the core of the plume. TKE is observed to be lower with increasing height across all cases. Within the gap, TKE is strongest at heights nearest to the heat patch.

Profiles of the TKE budget (Figure \ref{fig:All TKE Budget Profiles}m-r) show buoyancy is the dominant source of production, just as the 5 $\mathrm{m \cdot s^{-1}}$ cases. The homogeneous canopy exhibits a notable increase in buoyancy below 1h. The NC, EE, and IE cases show similar behavior in TKE budgets as the 5 $\mathrm{m \cdot s^{-1}}$ cases. However, shear production is observed to contribute more TKE in the gap canopies at 2h heights, while in the 5 $\mathrm{m \cdot s^{-1}}$ cases, shear production was weaker.

\section{Conclusion}

This study explored the role of canopy structure on buoyant plume dynamics using a suite of LES cases with horizontally heterogeneous canopy configurations. Six canopy setups - including no canopy (NC), external plume-edge canopy (EE), internal plume-edge canopy (IE), 200 m gap canopy (G2), 100 m gap canopy (G1), and homogeneous canopy (HC), were modeled with and without a surface heat flux patch of 5000 $\mathrm{W \cdot m^{-2}}$. Simulations were performed at prescribed geostrophic wind speeds of 0, 5, and 10 $\mathrm{m \cdot s^{-1}}$. Turbulence data from these experiments were analyzed for both mean and turbulent characteristics.

In the baseline (no plume) canopy flows, heterogeneous canopy configurations such as edge canopies (EC) and gap canopies (G2 and G1) developed recirculation zones downstream of the canopy edges. With the plume present, the overall flow characteristics were altered by the pressure gradient generated by the plume. At 0 $\mathrm{m \cdot s^{-1}}$ wind speed, variations in canopy configuration altered the plume updraft velocity. Pressure gradients above the heat patch varied by configuration; IE and HC showed a rapid change in pressure at the canopy top. The horizontal wind velocity on either side of the plume also differed with canopy structure. Large mean temperatures were observed at the canopy top for the IE and HC configurations.

At 5 $\mathrm{m \cdot s^{-1}}$, flow characteristics showed a significant reduction in horizontal velocity downstream of the plume, with pressure gradients differing by canopy configuration. Updraft speed and plume tilt varied accordingly. Gap canopy configurations exhibited recirculations within the gap, entrained toward the plume from the sides. Counter-rotating vortices formed downstream of the plume due to crosswind interaction. At 10 $\mathrm{m \cdot s^{-1}}$, downstream recirculation zones were absent, horizontal velocity reduction was less pronounced, the plume tilt angle became more shallow, and pressure gradients increased.

TKE budget analysis for the 0 $\mathrm{m \cdot s^{-1}}$ case showed that TKE was highest in the IE and HC configurations, with maximum values at approximately 5h in the vertical profiles. Buoyant production dominated over shear production, indicating turbulence generation was primarily driven by buoyancy effects in the absence of background wind. In the 5 $\mathrm{m \cdot s^{-1}}$ case, momentum flux profiles and spatial distributions indicated reductions centered over the plume source, with the gap canopies exhibiting the largest losses. TKE was greatest within the gap canopies, where peaks occurred near the canopy top. Gaps enhanced buoyant production, which remained the dominant contributor to turbulence, while shear production was also locally intensified within and above the gaps. In the 10 $\mathrm{m \cdot s^{-1}}$ cases, downward momentum flux was strongest, with the gap and homogeneous canopy configurations exhibiting the strongest downward momentum flux. TKE was also highest in these configurations, indicating enhanced turbulence generation. Buoyant production extended vertically above the homogeneous canopy, while shear production was further strengthened by the strong crosswind–plume interaction. This analysis improves our understanding of fire plume-vegetation-atmosphere interaction. Future studies involving fire behavior models could be used to compare against the baseline conditions provided in this study, and experiments could be devised for model-data comparison and validation.

\clearpage
\section*{Appendix: Additional Figures}

% \subsubsection{Baseline Canopy Flow (No Plume)}

\begin{figure}[ht!]
    \centering
    \begin{subfigure}{\textwidth}
        % \caption{}
        \includegraphics[width=\linewidth]{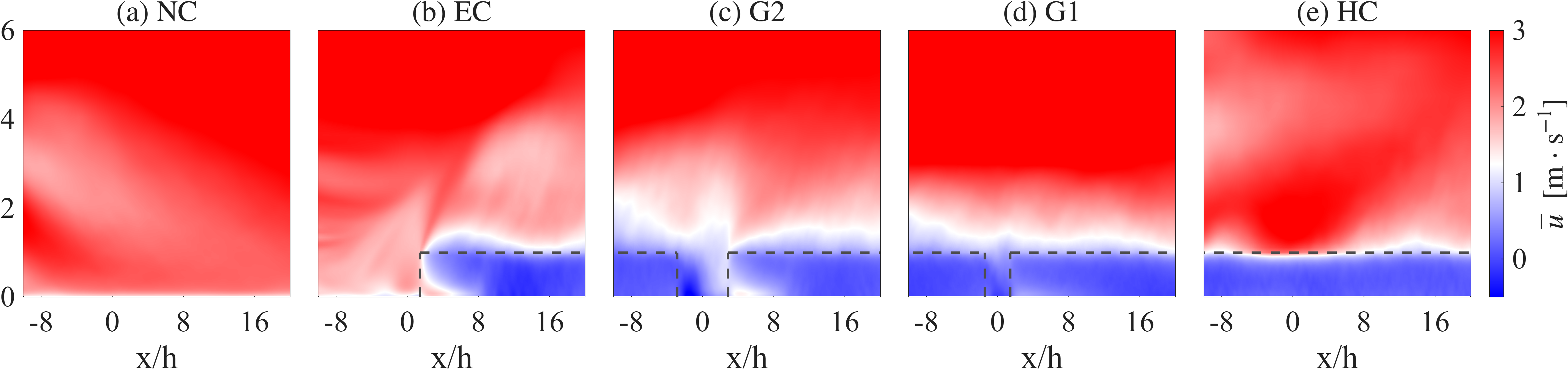}
    \end{subfigure}
    
    \vspace{0.5em}
    
    \begin{subfigure}{\textwidth}
        % \caption{}
        \includegraphics[width=\linewidth]{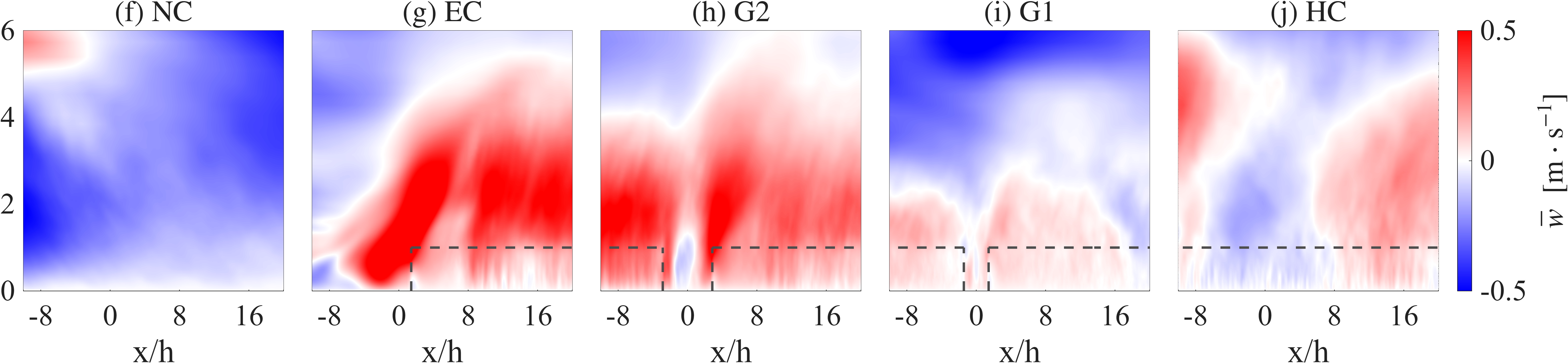}
    \end{subfigure}
    
    \vspace{0.5em}
    
    \begin{subfigure}{\textwidth}
        % \caption{}
        \includegraphics[width=\linewidth]{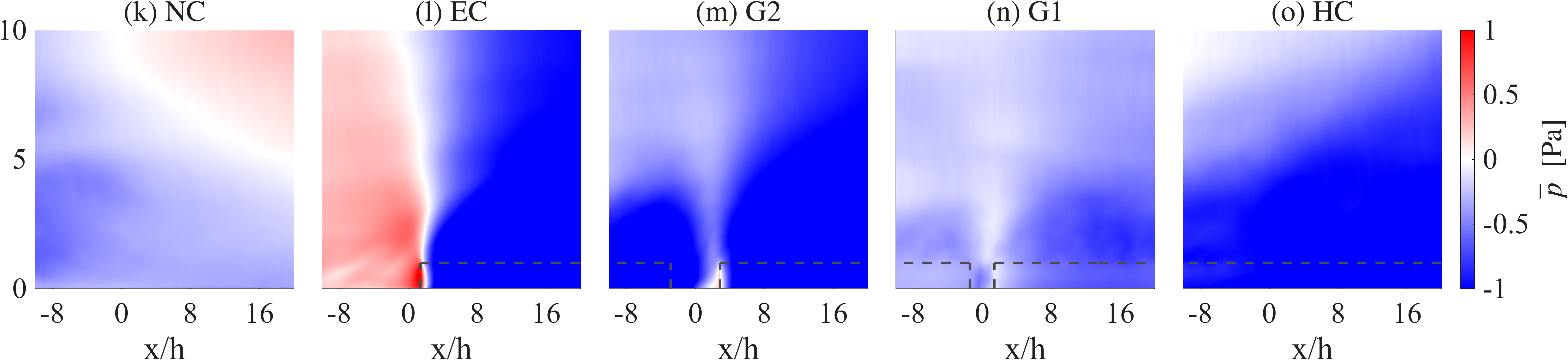}
    \end{subfigure}
    \caption{Spatial distribution of horizontal wind velocity (top row), vertical velocity (middle row), and pressure (bottom row) in the XZ-plane at $y = 500 \ \mathrm{m}$ for the $\mathrm{U = 5 \ m \cdot s^{-1}}$ no plume case for each canopy configuration. The x-axis origin is located at the center of the heat patch, and both the x and z axes are normalized by the canopy height ($h$).}
    \label{fig:U5 u no plume}
\end{figure}

% \subsection*{Mean Statistics}
% \subsubsection*{Plume in No Crosswind $(0 \ \mathrm{m \cdot s^{-1}})$}

\begin{figure}[ht!]
    \centering
    \includegraphics[width=\textwidth]{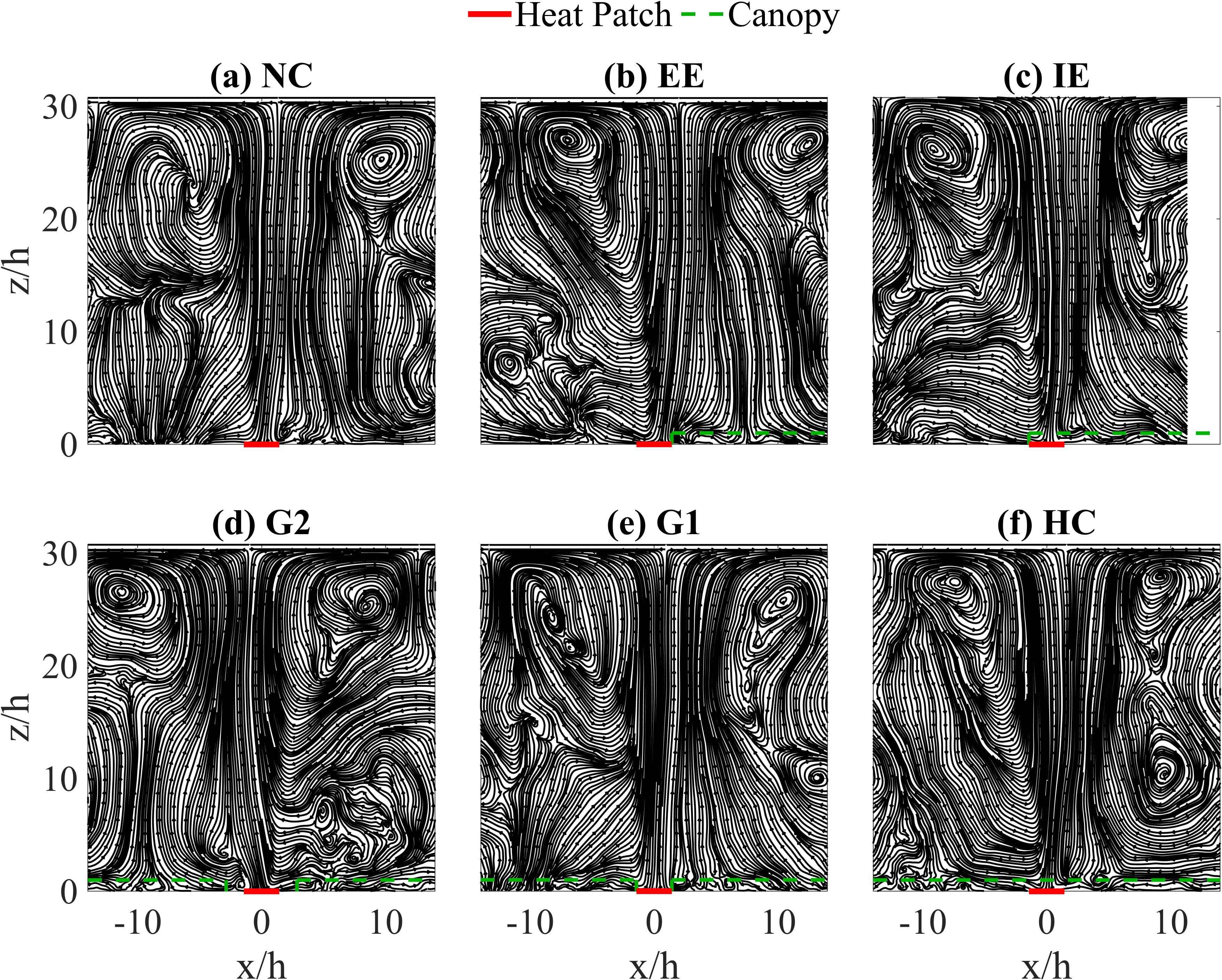}
    \caption{Zoomed out streamlines of no wind plume cases}
    \label{fig:U0 streamlines zoomed out}
\end{figure}

\begin{figure}[ht!]
    \centering
    \begin{subfigure}{0.25\textwidth}
        % \caption{}
        \includegraphics[width=\linewidth]{Figures/all_cases/spatial_plot_legend.png}
    \end{subfigure}
    
    \vspace{0.5em}
    
    \begin{subfigure}{\textwidth}
        % \caption{}
        \includegraphics[width=\linewidth]{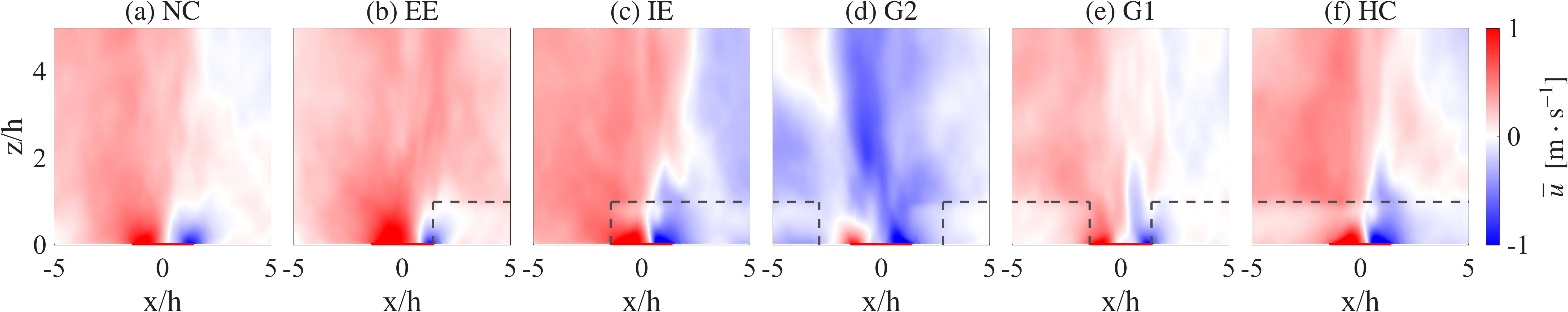}
    \end{subfigure}
    
    \vspace{0.5em}
    
    \begin{subfigure}{\textwidth}
        % \caption{}
        \includegraphics[width=\linewidth]{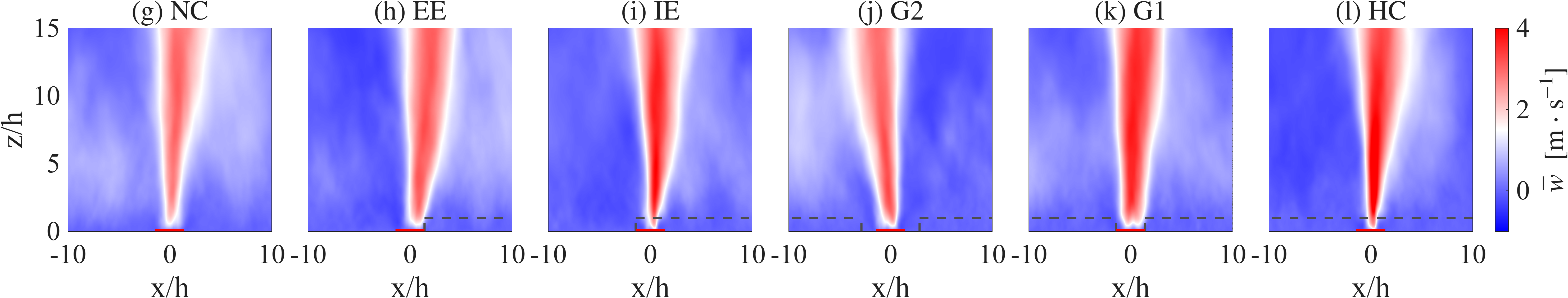}
    \end{subfigure}
    
    \vspace{0.5em}
    
    \begin{subfigure}{\textwidth}
        % \caption{}
        \includegraphics[width=\linewidth]{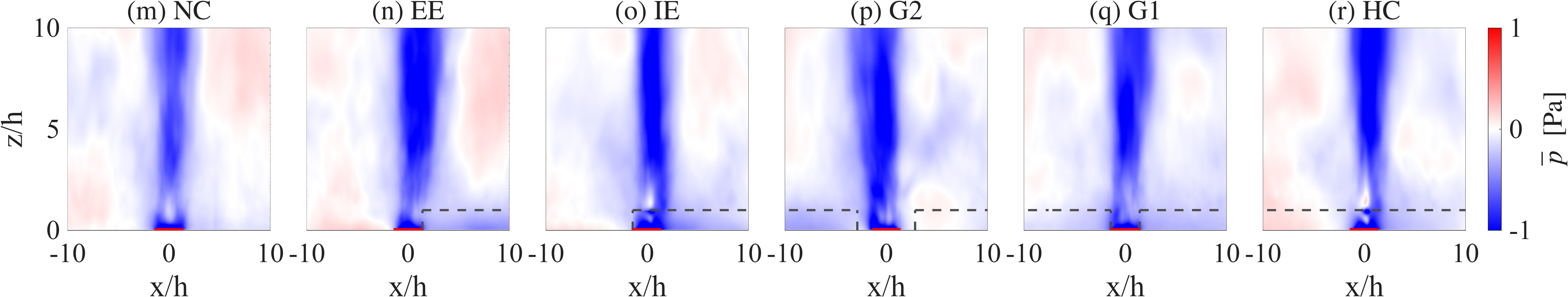}
    \end{subfigure}
    
    \vspace{0.5em}
    
    \begin{subfigure}{\textwidth}
        % \caption{}
        \includegraphics[width=\linewidth]{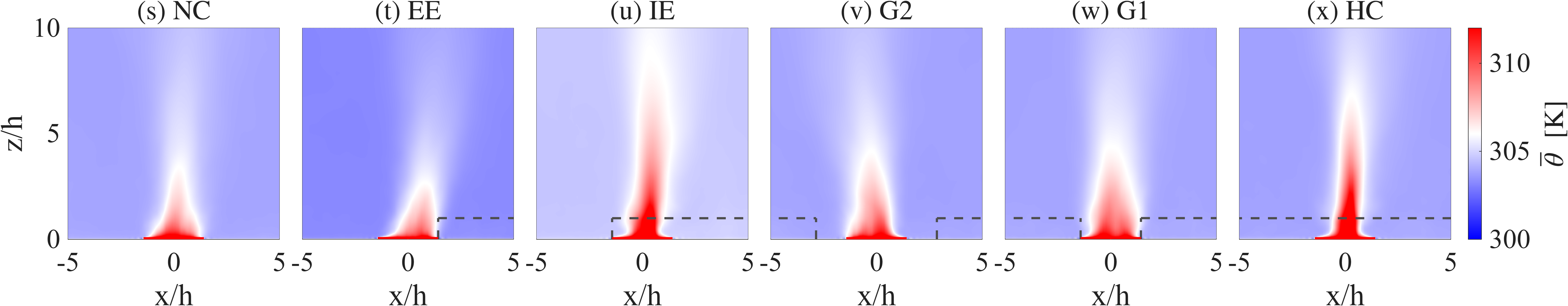}
    \end{subfigure}
    \caption{Spatial distribution of horizontal wind velocity (first row), vertical velocity (second row), pressure (third row), and temperature (fourth row) in the XZ-plane at $y = 500 \ \mathrm{m}$ for the $\mathrm{U = 0 \ m \cdot s^{-1}}$ wind speed case for each canopy configuration with plume. The x-axis origin is located at the center of the heat patch, and both the x and z axes are normalized by the canopy height ($h$).}
    \label{fig:U0 spatial mean variables}
\end{figure}

% \subsubsection{Plume in Crosswind $(5 \ \mathrm{m \cdot s^{-1}})$}

\begin{figure}[ht!]
    \centering
    \begin{subfigure}{0.25\textwidth}
        % \caption{}
        \includegraphics[width=\linewidth]{Figures/all_cases/spatial_plot_legend.png}
    \end{subfigure}

    \vspace{0.5em}

    \begin{subfigure}{\textwidth}
        % \caption{}
        \includegraphics[width=\linewidth]{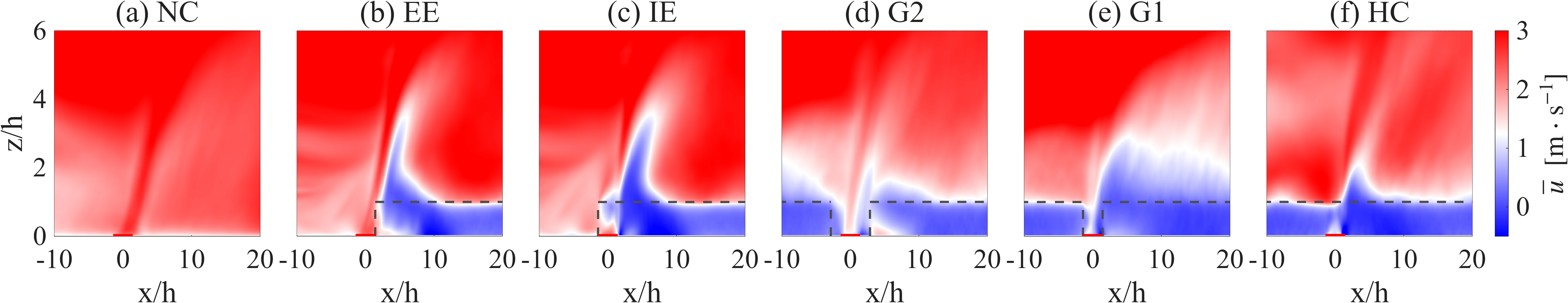}
    \end{subfigure}

    \vspace{0.5em}
    
    \begin{subfigure}{\textwidth}
        % \caption{}
        \includegraphics[width=\linewidth]{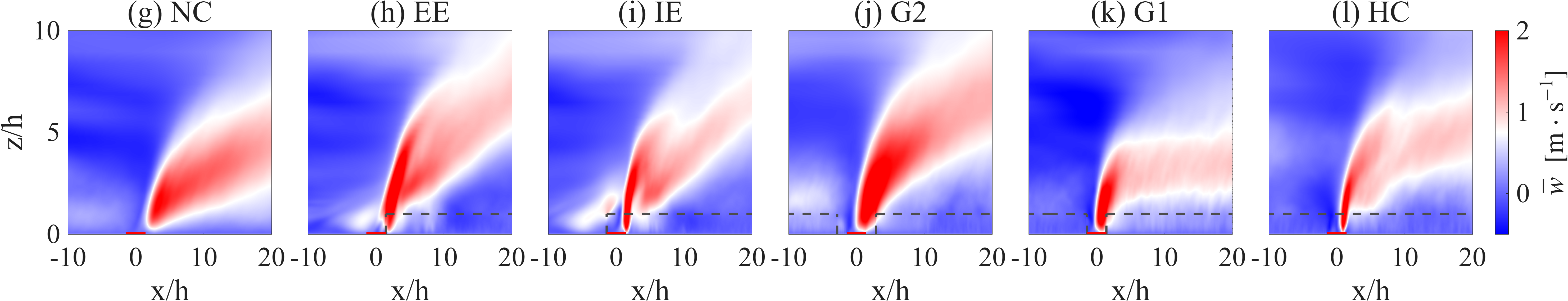}
    \end{subfigure}

    \vspace{0.5em}
    
    \begin{subfigure}{\textwidth}
        % \caption{}
        \includegraphics[width=\linewidth]{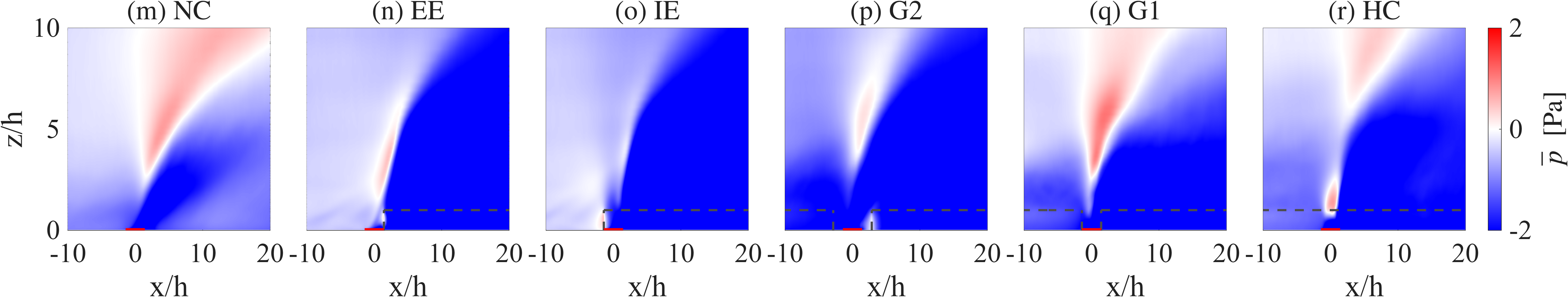}
    \end{subfigure}

    \vspace{0.5em}
    
    \begin{subfigure}{\textwidth}
        % \caption{}
        \includegraphics[width=\linewidth]{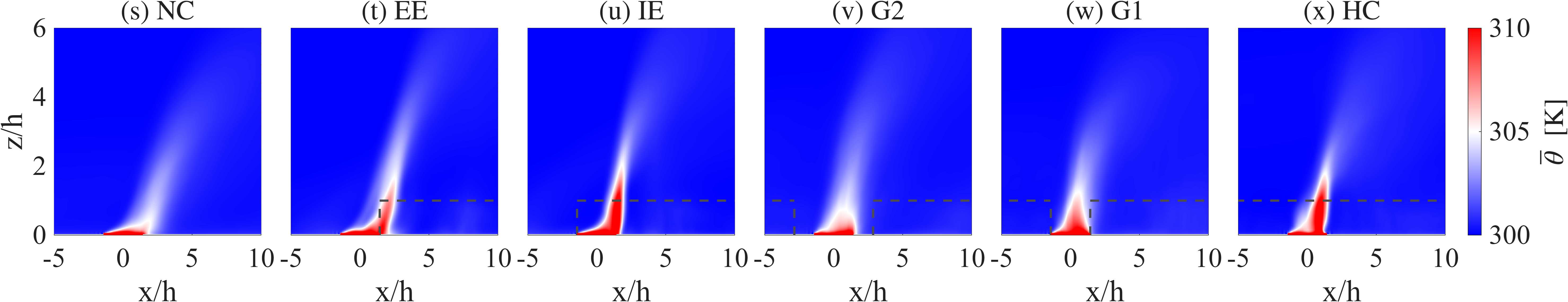}
    \end{subfigure}
    \caption{Spatial distribution of horizontal wind velocity (first row), vertical velocity (second row), pressure (third row), and temperature (fourth row) in the XZ-plane at $y = 500 \ \mathrm{m}$ for the $\mathrm{U = 5 \ m \cdot s^{-1}}$ wind speed case for each canopy configuration with plume. The x-axis origin is located at the center of the heat patch, and both the x and z axes are normalized by the canopy height ($h$).}
    \label{fig:U5 spatial mean variables}
\end{figure}

% \subsubsection{Plume in Crosswind $(10 \ \mathrm{m \cdot s^{-1}})$}

\begin{figure}[htbp]
    \centering
    \begin{subfigure}{0.75\textwidth}
        % \caption{}
        \includegraphics[width=\linewidth]{Figures/all_cases/legend.png}
        % \label{fig:U10 u vertical z/h=-3}
    \end{subfigure}
    \hfill
    \begin{subfigure}{\textwidth}
        % \caption{}
        \includegraphics[width=\linewidth]{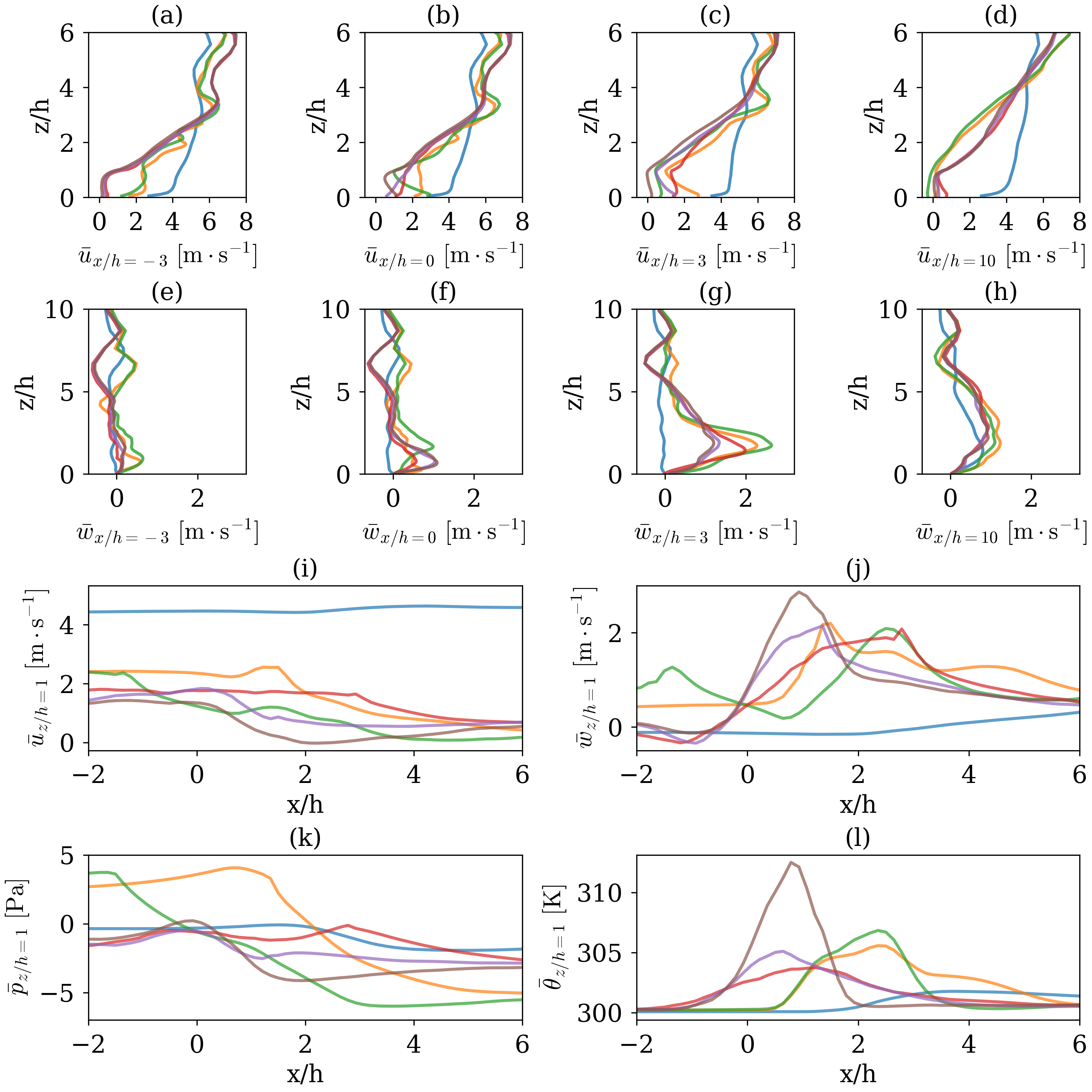}
        % \label{fig:U10 u vertical z/h=0}
    \end{subfigure}
    \caption{Profiles of mean flow variables for the crosswind case $(\mathrm{U=10 \ m \cdot s^{-1}})$. Vertical profiles of $\bar{u}$ (a-d) and $\bar{w}$ (e-h) at $x/h=-3,0,3, \ \text{and} \ 10$; horizontal profiles of $\bar{u}$ (i) and $\bar{w}$ at $z/h=1$ (j); horizontal profiles of $\bar{p}$ (k) and $\bar{\theta}$ at $z/h=1$ (l).}
    \label{fig:U10 mean variables}
\end{figure}

\begin{figure}[ht!]
    \centering
    \begin{subfigure}{0.25\textwidth}
        % \caption{}
        \includegraphics[width=\linewidth]{Figures/all_cases/spatial_plot_legend.png}
    \end{subfigure}
    
    \vspace{0.5em}
    
    \begin{subfigure}{\textwidth}
        % \caption{}
        \includegraphics[width=\linewidth]{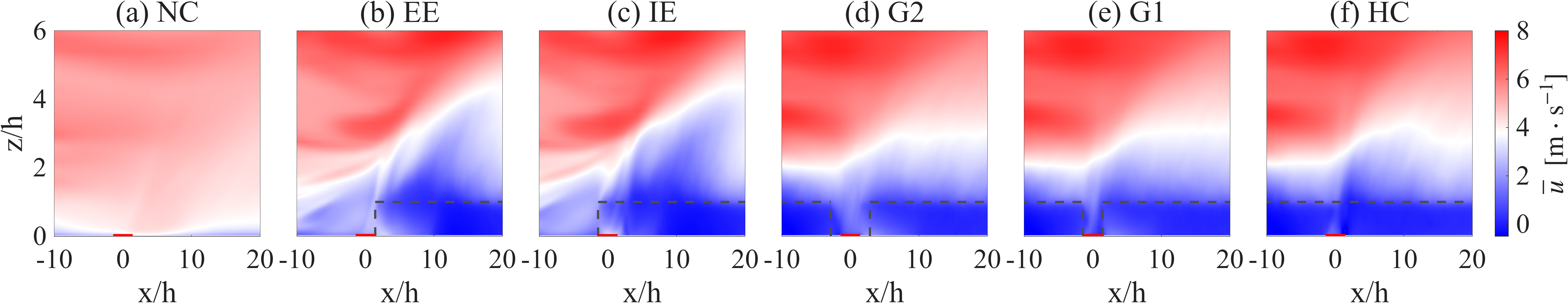}
    \end{subfigure}
    
    \vspace{0.5em}
    
    \begin{subfigure}{\textwidth}
        % \caption{}
        \includegraphics[width=\linewidth]{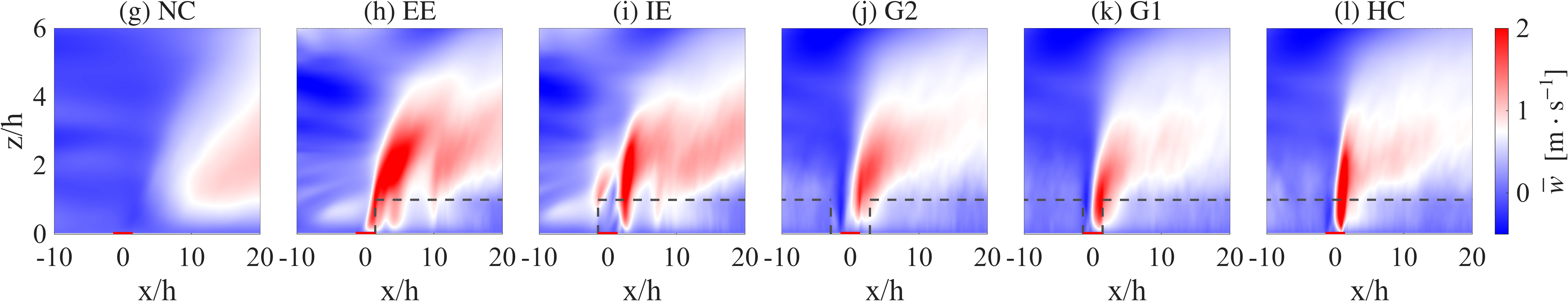}
    \end{subfigure}
    
    \vspace{0.5em}
    
    \begin{subfigure}{\textwidth}
        % \caption{}
        \includegraphics[width=\linewidth]{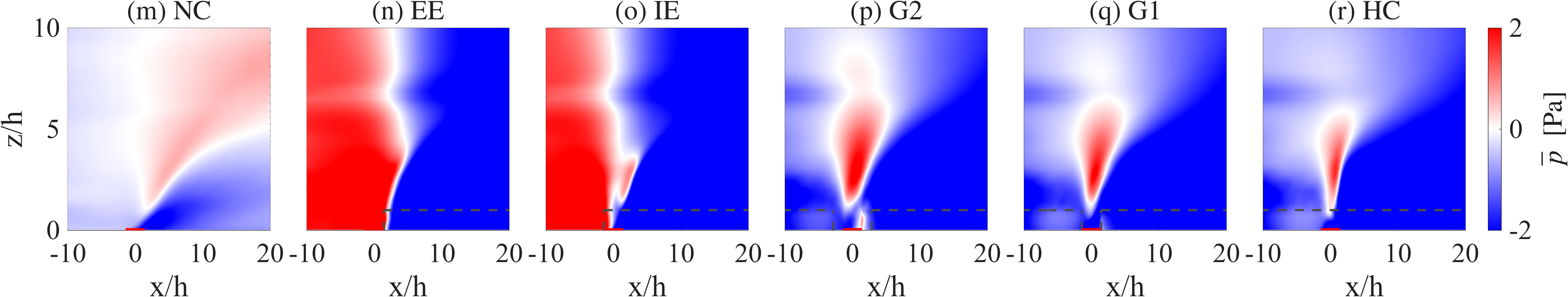}
    \end{subfigure}
    
    \vspace{0.5em}
    
    \begin{subfigure}{\textwidth}
        % \caption{}
        \includegraphics[width=\linewidth]{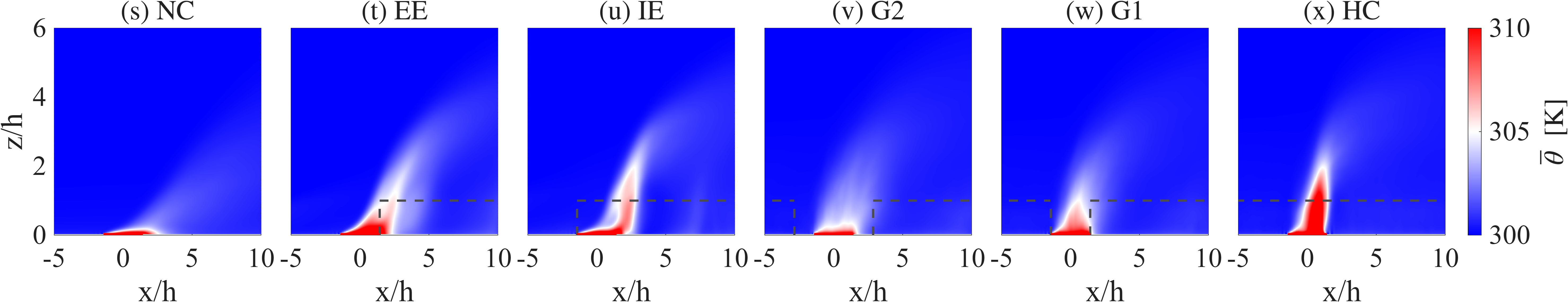}
    \end{subfigure}
    \caption{Spatial distribution of horizontal wind velocity (first row), vertical velocity (second row), pressure (third row), and temperature (fourth row) in the XZ-plane at $y = 500 \ \mathrm{m}$ for the $\mathrm{U = 10 \ m \cdot s^{-1}}$ wind speed case for each canopy configuration with plume. The x-axis origin is located at the center of the heat patch, and both the x and z axes are normalized by the canopy height ($h$).}
    \label{fig:U10 spatial mean variables}
\end{figure}

\begin{figure}[ht!]
    \centering
    \begin{subfigure}{0.25\textwidth}
        % \caption{}
        \includegraphics[width=\linewidth]{Figures/all_cases/spatial_plot_legend.png}
    \end{subfigure}
    
    \vspace{0.5em}
    
    \begin{subfigure}{\textwidth}
        % \caption{}
        \includegraphics[width=\linewidth]{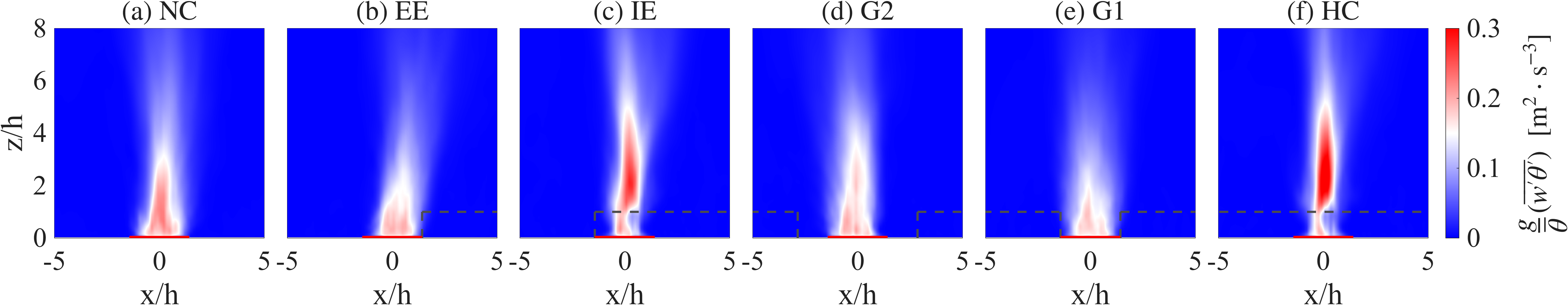}
    \end{subfigure}

    \vspace{0.5em}
    
    \begin{subfigure}{\textwidth}
        % \caption{}
        \includegraphics[width=\linewidth]{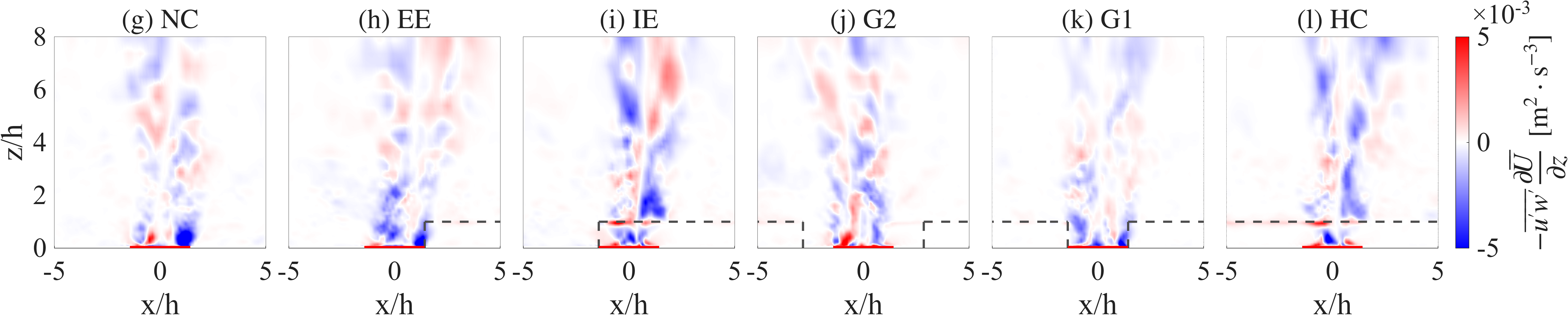}
    \end{subfigure}

    \vspace{0.5em}
    
    \begin{subfigure}{\textwidth}
        % \caption{}
        \includegraphics[width=\linewidth]{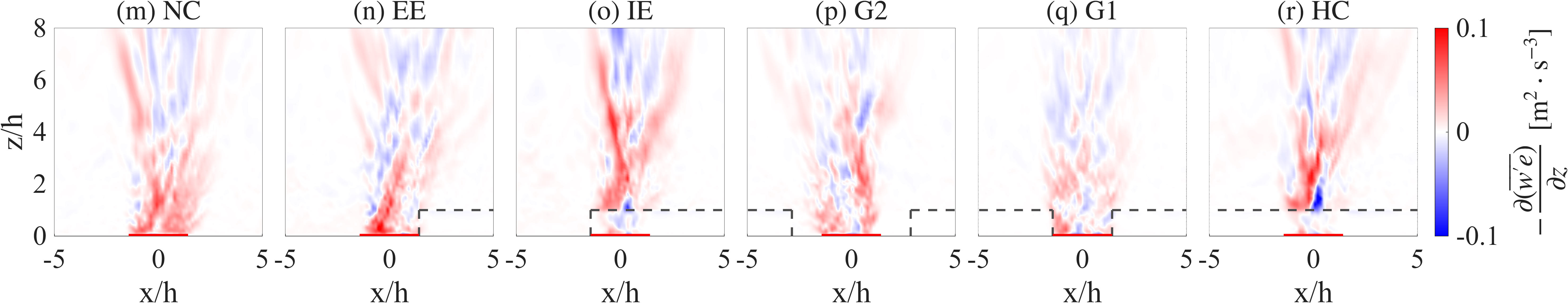}
    \end{subfigure}
    \caption{Spatial distribution of TKE budget terms in the XZ-plane at $y = 500 \ \mathrm{m}$ for the $\mathrm{U = 0 \ m \cdot s^{-1}}$ wind speed case of buoyancy (top row), shear (middle row), and transport (bottom row) for each canopy configuration with plume. The x-axis origin is located at the center of the heat patch, and both the x and z axes are normalized by the canopy height ($h$).}
    \label{fig:U0 TKE Budget spatial}
\end{figure}

\begin{figure}[ht!]
    \centering
    \begin{subfigure}{0.25\textwidth}
        % \caption{}
        \includegraphics[width=\linewidth]{Figures/all_cases/spatial_plot_legend.png}
    \end{subfigure}
    
    \vspace{0.5em}
    
    \begin{subfigure}{\textwidth}
        % \caption{}
        \includegraphics[width=\linewidth]{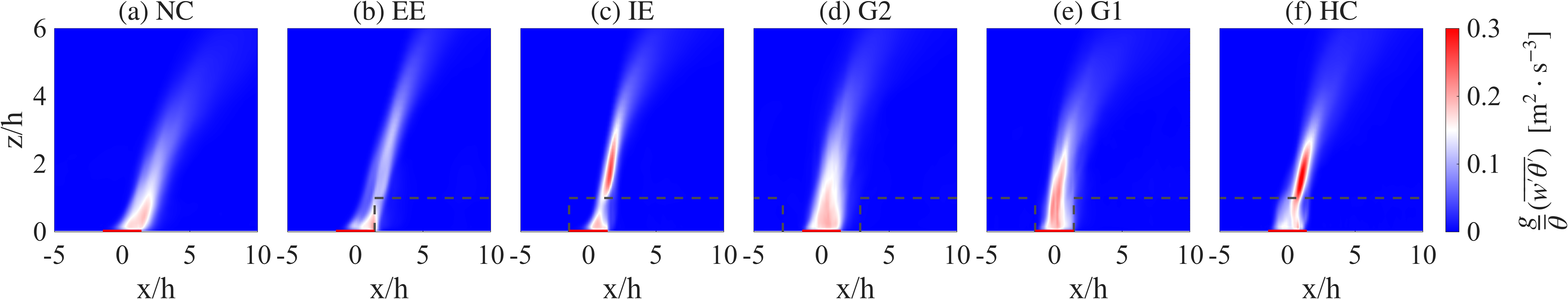}
    \end{subfigure}

    \vspace{0.5em}
    
    \begin{subfigure}{\textwidth}
        % \caption{}
        \includegraphics[width=\linewidth]{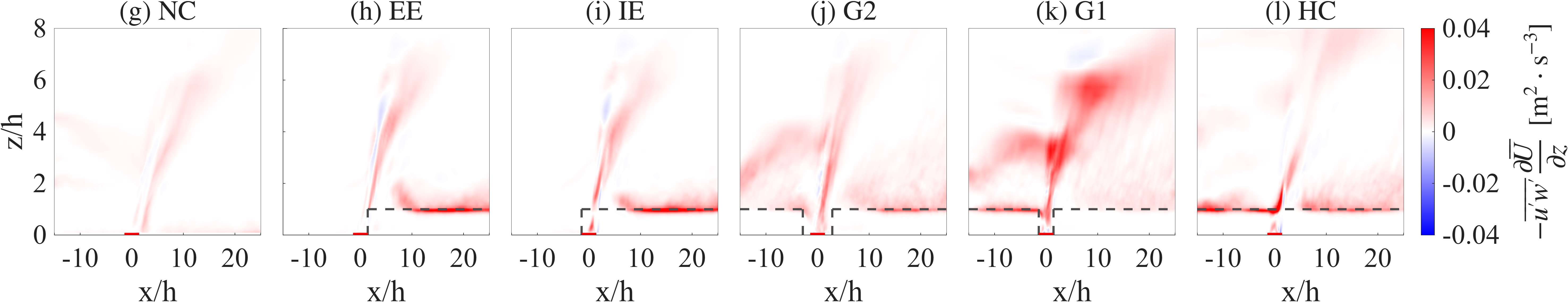}
    \end{subfigure}

    \vspace{0.5em}
    
    \begin{subfigure}{\textwidth}
        % \caption{}
        \includegraphics[width=\linewidth]{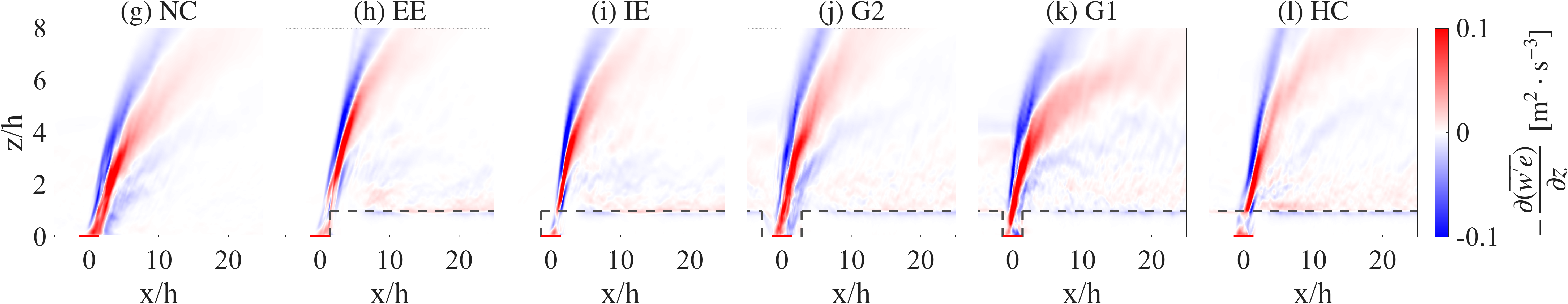}
    \end{subfigure}
    \caption{Spatial distribution of TKE budget terms in the XZ-plane at $y = 500 \ \mathrm{m}$ for the $\mathrm{U = 5 \ m \cdot s^{-1}}$ wind speed case of buoyancy (top row), shear (middle row), and transport (bottom row) for each canopy configuration with plume. The x-axis origin is located at the center of the heat patch, and both the x and z axes are normalized by the canopy height ($h$).}
    \label{fig:U5 TKE Budget spatial}
\end{figure}

\clearpage % forces references to be on new page
\bibliographystyle{spbasic_updated}     
\bibliography{ref} 

\end{document}